\documentclass[a4paper,12pt]{article}
\usepackage[english]{babel}
\usepackage{amsmath}
\usepackage{amssymb}
\usepackage{bbm}
\usepackage{color}
\usepackage{cite}
\usepackage{xcolor,hyperref}
\usepackage{graphicx}
\usepackage{xfrac}
\usepackage{tensor}
\usepackage{enumitem}
\usepackage{subfigure}
\usepackage[titletoc,toc,title]{appendix}
\usepackage[font=footnotesize,labelfont=bf]{caption}
\usepackage[lmargin=2.5cm, rmargin=2.5cm,tmargin=2.5cm,bmargin=3.5cm]{geometry}
\numberwithin{equation}{section}
\newcommand{\naw}[1]{\left(#1\right)}
\newcommand{\com}[1]{\left[#1\right]}
\newcommand{\modu}[1]{\left|#1\right|}
\newcommand{\poisson}[1]{\left\{#1\right\}}

\title{Group-theoretical analysis of quantum complexity:\\ the oscillator group case }

\author{K. Andrzejewski,$^1$\footnote{krzysztof.andrzejewski@uni.lodz.pl}\quad  K. Bolonek-Laso\'n,$^2$\footnote{katarzyna.bolonek@uni.lodz.pl}\quad and P. Kosi\'nski$^1$\footnote{piotr.kosinski@uni.lodz.pl} }
\date{%
	\small{	$^1$Faculty of Physics and Applied Informatics, University of Lodz,\\ 149/153 Pomorska St., 90-236 Lodz, Poland\\$^2$Faculty of Economics and Sociology, University of Lodz,\\ 41/43 Rewolucji 1905 St., 90-214 Lodz,  Poland}}

\begin{document}
	\maketitle
	
\begin{abstract}
Motivated by the recent rapid development of complexity theory applied to quantum mechanical processes we present the complete derivation of Nielsen's complexity of unitaries belonging to the representations of oscillator group. Our approach is based on the observation that the whole problem refers to the structure of the underlying group. The questions concerning the complexity of particular unitaries are solved by lifting the abstract structure to the operator level by considering the relevant unitary representation. For the class of right-invariant metrics obeying natural invariance condition we solve the geodesic equations on oscillator group. The solution is given explicitly in terms of elementary functions. Imposing the boundary conditions yield a transcendental equation and the length of the geodesic is given in terms of the solutions to the latter. Since the unitary irreducible representations of oscillator group are classified this allows us to compute, at least in principle, the complexity of any unitary operator belonging to the representation.	
\end{abstract}
\newpage	
\section{Introduction and conceptual insight}
\indent In recent years the notion of complexity has been playing an increasingly important role in the analysis of various physical processes. A particularly rewarding area for applying the concept of complexity is the physics of black holes and holography, where geometry, statistical physics, thermodynamics and information processing theory meet, inspiring each other.

Generally speaking, complexity is some measure of difficulty encountered in achieving a specific goal. It has a natural meaning in the context of computations. In particular, quantum complexity is originally defined in the theory of quantum computations. If one is interested in unitaries (the complexity of states may be viewed as a derived concept) the main observation is that any unitary operator (say, of unit determinant) in $d$-dimensional Hilbert space may be expressed as a product of unitary operators belonging to some set of universal gates. This set can be chosen to consist of single cubit and CNOT gates; if one is satisfied with approximate equality, only Hadamard, $\pi/8$ and CNOT gates are necessary \cite{NielsenChuang}. Once the set of universal gates is fixed, the complexity of a given unitary operator is defined as the minimal number of required gates to build it, exactly or approximately; in the latter case the complexity is sensitive to the level of accuracy.

Such a definition correctly captures the spirit of the actual computational process. It has, however, a serious drawback: determining the optimal circuit may be, in general, quite difficult task in the domain of discrete mathematics. Therefore, in the series of seminal papers Nielsen et al. \cite{Nielsen, NielsenDowling, Dowling} proposed to replace the discrete approach by a continuous one, based on differential geometry on Lie groups. This approach has been further considered, extended and applied in various contexts in numerous papers \cite{Jefferson,Brown,Yang,Khan,Hackl,Alves,Magan,Chapman,YangNiu,Brown1,Guo,Bhatta,YangAn,YangAn1,Camargo,Chapman1,Ali,Balasub,Doroudiani,Bhatta1,Bhatta2,Ghasemi,Chapman2,Pal,Pal1,Khorasani,Chowdhury,Pal2,Chowdhury1} (for review, see \cite{Baiguera}).

Let us summarize briefly the important points of Nielsen's proposal. One considers the system of $n$ qubits; the set of unitaries acting in $2^n$-dimensional Hilbert space of states forms defining (and, therefore, faithful) representation of $SU(2^n)$ group. The problem of determining the complexity of a given unitary operator can be now restated in purely group-theoretical terms. Namely, we select some subset of $SU(2^n)$, which plays the role of a set of unitary gates, such that any element of $SU(2^n)$ can be written as a group product of a finite number of elements of this subset; if we are only interested in approximate equalities, we can refer to the topology inherited from the norm topology of faithful representation (which is equivalent to the abstract topology defining $SU(2^n)$ as the Lie group). The complexity of a selected element is again defined as the minimal number of elements in such a product representation of this element. It is intuitively appealing to assume that the complexity, defined as above, is somehow related to the appropriately defined distance on $SU(2^n)$ group manifold. This intuition can be converted into a precise statement in a few steps. First, taking into account that all physical processes are described by the Schr\"odinger equation, $i\dot{U}U^{-1}=H$, one concludes that the relevant metric on the group manifold should be related to the right-invariant differential form $i\dot{U}U^{-1}$. It is important to note that this is differential on $SU(2^n)$ group manifold taking values in $sU(2^n)$ Lie algebra; its actual matrix form arises from replacing abstract $sU(2^n)$ generators by their representatives in faithful representation. Concluding, one deals with the right-invariant metric on the group manifold; the choice of right invariance is dictated by the conventions adopted in the standard formalism of quantum mechanics and the natural assumption that the real processing progresses toward positive time flow. It has been shown in \cite{Nielsen} that under some physically plausible assumptions the relevant metric is Finsler one. However, it seems sufficient to consider the special case of Riemannian geometry. Due to the right invariance such a metric is uniquely determined by choosing a positive definite quadratic form on the Lie algebra.

The next step is to select the set of universal gates. Since $SU(2^n)$ is connected any element can be written as a product of finite number of elements belonging to some neighbourhood of unity. However, not all elements of such a neighbourhood are necessary. Note that the Lie algebra $sU(2^n)$ is spanned by the generalized Pauli matrices. According to the result quoted above it is sufficient to choose as universal gates only the elements obtained (under exponential mapping) from the subspace of Lie algebra spanned by one- and two-qubit Pauli matrices. Now comes the important point. Under some mild assumption concerning the metric (cf. Theorem 1, Sec.~II, in \cite{Nielsen}) the distance between the unit element and a given one, defined as the minimal length of a curve connecting them, provides a lower bound on the complexity of the latter. This bound can be improved by the judicious choice of the metrics; due to the form of universal gates it should be chosen in such a way as to ``punish'' the directions in the group manifold generated by more than two-qubit Pauli operators.

The above described approach, devised by Nielsen and collaborators, is inspiring, mathematically elegant and allows to avoid complicated combinatorial reasonings. It works nicely in the case of finite-dimensional spaces of states. However, going beyond $n$-qubit case one encounters basic problems. The main point is that, since typically we are dealing with quantum systems described by infinite-dimensional space of states, the set of unitaries form an infinite-dimensional group.  Therefore, the tools from differential geometry on infinite-dimensional manifolds are necessary which makes things rather complicated. The natural question arises whether we can reformulate the complexity problem in a way that more modest mathematical tools are sufficient.

The aim of the present paper is to provide the arguments supporting this point of view. The main one is based on the observation that, although, in principle, all hermitean operation represent (modulo superselection rules) observables (while all unitaries represent admissible transformations/evolution), only a relatively small subset of these observables is physically relevant. We shall further dwell on this argument later on.

In order to explain our arguments in more detail let us look at the problem from a bird's eye view. To this end remind briefly basic characteristics of quantum description. As typically discussed in textbooks one starts with some classical dynamical system described within Hamiltonian formalism and applies the algorithm of the so-called canonical quantization to produce the quantum system which dynamics coincides with the initial one in macroscopic limit (obviously, such a procedure is by far not unique as one can always add terms proportional to positive powers of Planck's constant). However, this is turning the matter upside down. The correct description should be quantum-mechanical from the very beginning including the explanation of how the world looks like in the macroscopic limit. This is achieved as follows. One adopts the general postulates of quantum theory (superposition principle, linear space of states, linear transformations as observables, probabilistic interpretation) to set the stage. Then comes the important point: one selects the set of relevant symmetries. Typically, they are described by finite-dimensional Lie group (which can take the form of direct product of groups describing various symmetries). The choice of symmetry group defines the class of physical systems. Then, according to the general postulates of quantum mechanics, mentioned above, any particular system of this class is described by a specific unitary (in general, projective) representation of the symmetry group. Therefore, once the physically motivated choice of symmetry is made, all systems of a given class can be explicitly described provided the mathematical problem of finding all unitary representations of a given group is solved. For example, assuming space-time symmetry, non-relativistic or relativistic, one finds that all elementary systems as well as composite ones as a whole are described by unitary irreducible representations of Galilei or Poincare groups, respectively. Most interesting properties of elementary systems follow from the structure of these groups. Any elementary particle, whether it is an electron, Higgs particle or intermediate boson, is characterized by its mass and spin and this physical conclusion is dictated by the mathematics of space-time symmetry. Moreover, the multiparticle states are also uniquely described with the help of standard mathematical tool - the multiplication of representations. Obviously, further information concerning the form of interaction is necessary but is also based on the assumptions involving mainly symmetry arguments.

Generalizing, the crucial conclusion is that many (most?) physical properties of any class of physical systems can be described in terms of the structure of underlying abstract Lie group of symmetry transformations. On the other hand, the properties of the particular physical system belonging to a given class are related to the structure of the specific representation of symmetry group. This distinction between the physical properties of the class of systems (abstract Lie group level) and those of a particular member of this class (representation level) we find very important, also in the context of complexity analysis. 
There is a number of arguments supporting this conclusion. One of the most important is the following. As we have already mentioned almost any hermitean operator can be viewed as an observable. For example, one can consider $x^3p^2+p^2x^3$, however, its physical meaning is fairly unclear and rather uninteresting. On the other hand, there always exists a (relatively small) set of observables with clear physical interpretation, including energy, linear and angular momentum, charge, etc. Their importance stems from the fact that they are defined via quantum counterpart of Noether theorem. Namely, as in the classical Hamiltonian formalism, they are generators of symmetry transformations, i.e. they correspond to the elements of the Lie algebra of the symmetry group. Things get slightly more complicated if one considers more general forms of symmetries-spontaneously or explicitly broken ones and gauge symmetries but, essentially, the relevant observables are related to the symmetry transformations. Summarizing, one can say that the observables with clear physical interpretation are the elements of (universal enveloping of) Lie algebra of $G$ while physically relevant transformations - the elements of symmetry group itself. 

Summing up the above the following approach to the complexity problem seems to be reasonable. Given a class of physical systems characterized by a symmetry described by a (connected) Lie group $G$ one can identify physically relevant transformations/evolution operators as the elements of $G$; they are subject to the complexity analysis. On the other hand, the universal gates should also be unitaries with clear physical interpretation, i.e. the ones represented by the elements of $G$. This implies that the set of universal gates can be defined as subset $\Gamma\subset G$ such that its elements generate the whole group $G$ in form of finite products of elements of $\Gamma$; the complexity of a given element of $G$ is defined as a minimal number of factors in such a product. Then the problem of computing the complexity can be solved on the abstract group level without any reference to a specific representation. Within this setting the Theorem 1, Sec.~II, in \cite{Nielsen} is essentially a group-theoretical one and can be proven without referring to a particular representation of the group. 

Two points are worth to be stressed:
\begin{itemize}
	\item[-] one avoids entering into complicated domain of infinite-dimensional geometry by realizing that among huge variety of formally admissible observables and transformations only those related to symmetries are physically relevant; due to this only the latter are interesting from the point of view of complexity theory (including the construction of universal gates);
	\item[-] in the $n$-qubit case one considers the set of unitaries in $2^n$-dimensional space; the latter can be viewed both as abstract $U(2^n)$ group as well as its faithful representation; therefore, the important distinction we stressed above, is less transparent. 
\end{itemize}
The approach described here has several advantages:
\begin{itemize}
	\item[-] several authors analyze the geodesic equations referring to a particular representation. This suggests the use of solving methods typical for quantum mechanics, in particular - the Dyson expansion. With this tool the symmetry structure becomes somewhat obscure and, as a result, one has to be satisfied with low order contributions which typically do not reflect properly global properties of the solutions. On the contrary, ``abstract'' approach leads immediately to the Hamiltonian system with high degree of symmetry making tools of Hamiltonian dynamics (integrals of motion, momentum maps, Hamiltonian reduction etc.) immediately available;
	\item[-] an important step in Nielsen's approach is the proper choice of right-invariant metric on group manifold. As it has been mentioned above the metric must be chosen in such a way as to ``punish'' the directions in group manifold which are not related to the universal gates. In fact, the primary motivation for ``continuous'' approach is to provide a reasonable estimate for the complexity defined in terms of gates. The latter may (and, in general, does) depend not only on the symmetry assumed but also on the choice of particular system in the symmetry class. Therefore, one has to choose the metric carefully to reflect the structure of the particular set of universal gates. This requires flexibility in the choice of the relevant metric. Again, in the examples encountered in the literature, the choice of metric is often dictated by the simplicity of the resulting geodesic equations rather than by the above criterion. Again, the approach advocated here allows for considering much wider class of metrics. With more exhaustive knowledge about the geodesics for the set of physically reasonable right-invariant metrics one obtains effective tool for computing the complexity for various physical systems;
	\item[-] some authors argue that one should use a bi-invariant metric instead of the right-invariant one \cite{YangAn,YangAn1}. This is because quantum mechanical description is defined up to an unitary equivalence while right-invariant metric is generally, not left-invariant. However, we should keep in mind that unitary transformations (unitary equivalence) concern not only unitary operators (evolution operators) under consideration but also the universal gates the complexity definition refers to. For example, in the $n$-qubit case the basis is fixed by demanding that the tensor product structure is explicit in this basis. Therefore, performing a similarity transformation on elements of $SU(2^n)$ but computing the complexity with respect to the old universal gates is not the equivalent description of the same physical situation. On the other hand, applying the similarity transformation to the gates only implies the change of metric. 
\end{itemize}   	

In the present paper we discuss in some detail one example of symmetry group: the oscillator group \cite{Streater}. Our aim is to show explicitly how the ideas exposed above work. We have chosen oscillator group because of (\textit{i}) its physical relevance and (\textit{ii}) simplicity of the corresponding mathematics. In Section \ref{sec2} we describe in some detail the formal properties of oscillator group, including its global structure, automorphisms, non-exponential character and unitary irreducible representations. Section \ref{sec3} is devoted to the analysis of geodesic equations for three-parameter family of right-invariant metrics. We stress the integrability properties and show that the geodesics may be viewed as trajectories of charged particles in uniform magnetic field. We stress that, generally, we are dealing with the whole families of geodesics and the choice of the shortest one is by far not straightforward. In Section \ref{sec4} we apply the results obtained to compute the complexity of certain unitaries with transparent physical interpretation. Finally, Section \ref{sec5} is devoted to some conclusions and outlook.

\section{The oscillator group}\label{sec2}

Let us illustrate the general ideas presented above with the example of oscillator group. It is closely related to the dynamics of harmonic oscillator and therefore underlies the symmetry structure of numerous physical systems. Moreover, it admits infinite dimensional irreducible representations.
We describe below some necessary information about the oscillator group and algebra we shall need in what follows. Let us start with the four-dimensional Lie algebra, spanned by the generators $E$, $Q$, $P$ and $H$, obeying
\begin{align}
	& \com{Q,P}=iE, \label{a1}\\
	& \com{Q,H}=iP, \label{a2}\\
	& \com{P,H}=-iQ, \label{a3}\\
	& \com{E,\cdot}=0.\label{a4}
	\end{align}
The oscillator group $G$ is heuristically defined by considering the formal products of the form
\begin{equation}
	g(e,\alpha,q,p)\equiv e^{ieE}e^{i\alpha H}e^{i(pQ+qP)};\label{a5}
\end{equation}
we choose such a form because we cannot expect a priori that $G$ is exponential (actually, it is not) while the Heisenberg-Weyl subgroup, spanned by $E$, $Q$ and $P$, being nilpotent, is exponential. The composition law, inferred from \eqref{a1}$\div$\eqref{a5} reads:
\begin{equation}
	g(e,\alpha,q,p)\cdot g(e',\alpha',q',p')=g(e'',\alpha'',q'',p''),\label{a6}
\end{equation}
where
\begin{align}
& e''=e+e'+\frac{1}{2}p'(q\cos\alpha'-p\sin\alpha')-\frac{1}{2}q'(p\cos\alpha'+q\sin\alpha'),\label{a7}\\
& q''=q'+q\cos\alpha'-p\sin\alpha',\label{a8}\\
& p''=p'+p\cos\alpha'+q\sin\alpha',\label{a9}\\
& \alpha''=\alpha+\alpha'.\label{a10}
\end{align}
$G$ is defined as topologically $\mathbbm{R}^4$ equipped with the composition law \eqref{a6}$\div$\eqref{a10}. It has the form
\begin{equation}
	G=\mathbbm{R}\ltimes G_H, \label{a11}
	\end{equation}
where $\mathbbm{R}$ is viewed as an abelian group with addition as group multiplication, generated by $H$, while $G_H$ is the Heisenberg-Weyl group generated by $E$, $Q$ and $P$.

$G$ is not exponential \cite{Streater}. To see this, let us select an element of Lie algebra
\begin{equation}
	X=x^eE+x^pQ+x^qP+x^\alpha H, \label{a12}
\end{equation}
 and define
 \begin{equation}
 	g(t)\equiv e^{itX}. 
 \label{a13}
 	\end{equation}
The exponential mapping $(x^e,x^q,x^p,x^\alpha)\rightarrow (e,\alpha,q,p)$ is given by $g(1)$. Eq.~\eqref{a13} implies
\begin{equation}
\frac{\mathrm{d}g}{\mathrm{d}t}=iXg=igX, 
\end{equation}
or, in terms of coordinates,
\begin{equation}
	\dot{a}^i={{\mu_L}^i}_j(a)x^j={{\mu_R}^i}_j(a)x^j, \label{a15}                                                                                     
\end{equation}
with $(a^i)=(e,\alpha,q,p)$, $(x^i)=(x^e,x^q,x^p,x^\alpha)$ and $\mu_{L,R}$ are the components of left- and right-invariant vector fields (cf.~Appendix \ref{appendA}). Using eq.~\eqref{Ap8} from the Appendix one finds
\begin{alignat}{4}
&{{\mu_L}^e}_e=1,&\qquad &{{\mu_L}^q}_e=0,&\qquad &{{\mu_L}^p}_e=0,&\qquad &{{\mu_L}^\alpha}_e=0,\label{a16}\\
&{{\mu_L}^e}_q=-\frac{1}{2}p,&\qquad  &{{\mu_L}^q}_q=1, &\qquad &{{\mu_L}^p}_q=0, &\qquad &{{\mu_L}^\alpha}_q=0,\label{a17}\\
&{{\mu_L}^e}_p=\frac{1}{2}q,&\qquad & {{\mu_L}^q}_p=0, &\qquad &{{\mu_L}^p}_p=1,&\qquad & {{\mu_L}^\alpha}_p=0,\label{a18}	\\
&{{\mu_L}^e}_\alpha=0,&\qquad &{{\mu_L}^q}_\alpha=-p, & \qquad &{{\mu_L}^p}_\alpha=q,&\qquad & {{\mu_L}^\alpha}_\alpha=1.\label{a19}
\end{alignat}

We have written out the components of left-invariant vector fields since this form of eq.~\eqref{a15} is more convenient with the parametrization \eqref{a5} adopted here. Eqs.~\eqref{a15} read then
\begin{align}
	& \dot{e}=x^e-\frac{1}{2}x^q p+\frac{1}{2}x^p q, \label{a20} \\
	&\dot{q}=x^q-x^\alpha p, \label{a21}\\
	&\dot{p}=x^p+x^\alpha q, \label{a22}\\
	&\dot{\alpha}=x^\alpha.
\end{align}
Solving the above set, putting $t=1$ and inverting the result with respect to $x$'s one obtains:
\begin{align}
	&x^e=e+\frac{1}{4}(q^2+p^2)\naw{\frac{\alpha-\sin\alpha}{1-\cos\alpha}},\label{a24}\\
	& x^q=\frac{\alpha}{2}\naw{p+\frac{q\sin\alpha}{1-\cos\alpha}}, \label{a25}\\
	& x^p=\frac{\alpha}{2}\naw{-q+\frac{p\sin\alpha}{1-\cos\alpha}},\label{a26}\\
	& x^\alpha=\alpha.\label{a27}
\end{align}
We see that the exponential mapping covers $G$ except the set of punctured hyperplanes $\alpha=2k\pi$, $k=\pm 1,\pm 2,\ldots$, $q^2+p^2\neq 0$.

Next, let us discuss the automorphisms of the algebra \eqref{a1}$\div$\eqref{a4}. It is straightforward to check that there exist two families of automorphisms:
\begin{align}
	& Q'=\mu Q+\nu P+(\nu\sigma+\mu\rho)E,\label{a28}\\
	& P'=-\nu Q+\mu P+(\mu\sigma-\nu\rho)E,\qquad \mu^2+\nu^2\neq 0,\label{a29}\\
	& H'=H+\rho Q+\sigma P+\tau E,\label{a30}\\
	& E'=(\mu^2+\nu^2)E,\label{a31}
\end{align}
or
\begin{align}
	& Q'=\mu Q+\nu P-(\nu\sigma+\mu\rho)E,\label{a32}\\
	& P'=\nu Q-\mu P+(\mu\sigma-\nu\rho)E,\label{a33}\\
	& H'=-H+\rho Q+\sigma P+\tau E,\label{a34}\\
	& E'=-(\mu^2+\nu^2)E.\label{a35}
\end{align}
Defining new parameters $e,\,q,\,p,\,\alpha$ by 
\begin{equation}
	e'E'+p'Q'+q'P'+\alpha'H'=eE+pQ+qP+\alpha H,
\end{equation}
one finds
\begin{align}
&e=(\mu^2+\nu^2)e'+(\nu\sigma+\mu\rho)p'+(\mu\sigma-\nu\rho)q'+\tau\alpha',\\
& p=\mu p'-\nu q'+\rho\alpha',\\
& q=\nu p'+\mu q'+\sigma\alpha',\\
&\alpha=\alpha',
\end{align}
and for the second family
\begin{align}
	&e=-(\mu^2+\nu^2)e'-(\nu\sigma+\mu\rho)p'+(\mu\sigma-\nu\rho)q'+\tau\alpha',\\
	& p=\mu p'+\nu q'+\rho\alpha',\\
	& q=\nu p'-\mu q'+\sigma\alpha',\\
	&\alpha=-\alpha'.
\end{align}
Once $G$ is defined one can look for its unitary irreducible representations. This is crucial from the complexity point of view. To this end note that
\begin{align}
	&C_1\equiv E,\label{a45}\\
	& C_2\equiv HE-\frac{1}{2}(Q^2+P^2),\label{a46}
\end{align}
commute with all generators (Casimir operators). Therefore, within an irreducible representation they are proportional to identity
\begin{align}
	&C_1=\omega I,\label{a47}\\
	&C_2=hI. \label{a48}
\end{align}
There are two, qualitatively different, cases. If $\omega =0$ we are dealing with homomorphic representation of $G$; actually, it is a representation of universal covering of two-dimensional Euclidean group $E(2)$. Its representations can be obtained by induction procedure, much like those of the  Poincare group. In what follows we will be interested in the case $\omega\neq 0$. In this case the center of the Heisenberg-Weyl subgroup, spanned by $Q$, $P$ and $E$, acts nontrivially. Then, by the Stone-von Neumann theorem there exists exactly one (up to an unitary equivalence) irreducible representation of the Heisenberg subgroup for any $\omega\neq 0$; it can be explicitly described, for instance, in terms of the Fock space.

Once the representation of Heisenberg-Weyl subgroup is selected, the representation of the oscillator group is fixed uniquely. In fact, it follows from \eqref{a46} and \eqref{a48} that
\begin{equation}
	H=\frac{h}{\omega}I+\frac{1}{2\omega}(Q^2+P^2).\label{a49}
\end{equation}
Due to \eqref{a47} the operators $\tilde{Q}\equiv Q/\omega$, $\tilde{P}=P$ form a canonical pair and
\begin{equation}
	H=\frac{h}{\omega}I+\frac{1}{2\omega}(\omega^2\tilde{Q}^2+\tilde{P}^2).\label{a50}
\end{equation}
Therefore, the spectrum of $H$ reads:
\begin{equation}
	h_n=\frac{h}{\omega}+\frac{\modu{\omega}}{\omega}\naw{n+\frac{1}{2}},\qquad n=0,1,2,\ldots
\end{equation}
We have to distinguish two cases. If $\frac{h}{\omega}+\frac{\modu{\omega}}{2\omega}$ is irrational, we are dealing with faithful representation of the subgroup generated by $H$. On the other hand, if
\begin{equation}
	\frac{h}{\omega}+\frac{\modu{\omega}}{2\omega}=\frac{k}{l}\mod 1,
\end{equation}
the kernel of the representation of this subgroup is the subgroup consisting of the powers of $\exp(2\pi liH)$. We obtain the faithful representation of $G/S$ where $S$ is the group of addition modulo $2\pi l$.

With such more detailed insight into the structure of oscillator group we can discuss the properties of the geodesics defined with respect to the right-invariant metrics on group manifold. This will be done in the next section.

\section{Geodesics}\label{sec3}
\vspace{-0.6 cm}
\subsection{Geodesic equations and their solutions}
Below we study the geodesics for the oscillator group that are relevant to the geometric approach to complexity, proposed in \cite{NielsenChuang,Nielsen,NielsenDowling}. In order to define the right-invariant metric on group manifold one has to select first the bi-linear form on the Lie algebra, $\eta_{ij}$, $i,j=e,\,q,\,p,\,\alpha$. The choice of $\eta_{ij}$ is, in principle, arbitrary; $\eta_{ij}$ may be any symmetric positive definite matrix. Let us note that the particular automorphism belonging to the family \eqref{a32}$\div$\eqref{a35}:
\begin{equation}
	Q\rightarrow P,\quad P\rightarrow-Q,\quad H\rightarrow H,\quad E\rightarrow E, \label{a53}
\end{equation}
corresponds, on the physical level, to the statement that the choice which variables, out of two canonically conjugated, correspond to coordinate and momentum, is arbitrary. Such an assumption is quite natural from the physical point of view. Therefore, we will assume that the metric is invariant under \eqref{a53}. Moreover, the geodesic equations are invariant under rescaling, $\eta_{ij}\rightarrow \lambda\eta_{ij}$, $\lambda\in\mathbbm{R}_+$. The invariance under \eqref{a53}, together with appropriate rescaling, implies the following form of $\eta$:
\begin{equation}
	\eta=\begin{pmatrix}
		a & 0 & 0 & b\\
		0 & 1 & 0 & 0\\
		0 & 0 & 1 & 0\\
		b & 0 & 0 & d\end{pmatrix},\qquad ad-b^2>0.\label{a54}
\end{equation}
It appears that this metric enjoys richer symmetry consisting of rotations in $q-p$ plane (the automorphisms \eqref{a28}$\div$\eqref{a31} with $\mu^2+\nu^2=1$, $\rho=\sigma=\tau=0$). This is seen in eqs.~\eqref{a85} and \eqref{a86} below where the length of geodesic depends on $q^2+p^2$ only. 
Once the metric of the tangent space to the unit element is defined it can be extended to the whole group manifold by the condition of right invariance.

One way to study the geodesics on $G$ is to work with its explicit realization (given by an unitary representation) \cite{Chowdhury,Chowdhury1}. Then we are dealing with the elements of $G$ (or some quotient group if the kernel of the representation is nontrivial) taking the form of unitary operators $U(g)$. In particular, the abstract equations \eqref{Ap5}, \eqref{Ap7} take the form
\begin{equation}
	i\frac{\mathrm{d}U}{\mathrm{d}t}U^+={{\lambda_R}^j}_k\dot{a}^k X_j\equiv\Pi^jX_j,\label{a55}
\end{equation} 
the only difference is that now $X_j$ are the generators of $G$ in a given representation. Once $\Pi^j(t)$ are known one can rewrite eq.~\eqref{a55} in the form of the Schr\"odinger equation:
\begin{equation}
	i\frac{\mathrm{d}U}{\mathrm{d}t}=\mathcal{H}U, \label{a56}
\end{equation}
with
\begin{equation}
\mathcal{H}(t)\equiv \Pi^i(t)X_i, \label{a57}
\end{equation}
being a hermitean Hamiltonian. Eq.~\eqref{a56} can be solved, at least approximately, using known techniques - the Dyson chronological product, Magnus expansion, perturbation theory etc. 

However, our problem is purely geometric and refers to the geometry of the group manifold itself. Therefore, we can deal directly with the equations \eqref{Ap34}:
\begin{equation}
	\frac{\mathrm{d}a^i}{\mathrm{d}t}={{\mu_R}^i}_j(a)\Pi^j(t). \label{a58}
\end{equation}
Once the geodesic $g=g(a(t))$ is determined one can consider the corresponding path $U(t)=U(g(a(t)))$ in the algebra of operators acting in the space of states.

The only additional information brought by the choice of representation concerns its kernel $K$. The representation reflects faithfully the structure of the quotient group $G/K$ and one should classify the geodesics on the latter. If $K$ is discrete this amounts only for the modification of topology.

Going back to our oscillator group we note that the Euler-Arnold equations, determining $\Pi^j(t)$ (cf.~eq.~\eqref{Ap33}), are integrable for $\eta$ given by eq.~\eqref{a54}. In fact, they take the form:
\begin{align}
	& a\frac{\mathrm{d}\Pi^e}{\mathrm{d}t}+b\frac{\mathrm{d}\Pi^\alpha}{\mathrm{d}t}=0,\label{a59}\\
	& b\frac{\mathrm{d}\Pi^e}{\mathrm{d}t}+d\frac{\mathrm{d}\Pi^\alpha}{\mathrm{d}t}=0,\label{a60}\\
	& \frac{\mathrm{d}\Pi^q}{\mathrm{d}t}=-(1+b)\Pi^\alpha\Pi^p-a\Pi^e\Pi^p,\label{a61}\\
	& \frac{\mathrm{d}\Pi^p}{\mathrm{d}t}=(1+b)\Pi^\alpha\Pi^q+a\Pi^e\Pi^q, \label{a62}
\end{align}
 with the solutions
 \begin{align}
 	& \Pi^e=A,\label{a63}\\
 	& \Pi^\alpha =B,\label{a64}\\
 	& \Pi^q=-D\sin(\nu t)+F\cos(\nu t),\label{a65}\\
 	& \Pi^p=D\cos(\nu t)+F\sin(\nu t), \label{a66}
 \end{align} 
where
\begin{equation}
	\nu\equiv aA+(b+1)B.\label{a67}
\end{equation}
Once $\Pi$'s are known one can refer to eqs.~\eqref{a58} in order to compute the coordinates of group elements. To this end we determine ${{\mu_R}^i}_j(a)$ from eq.~\eqref{Ap9}. We find
 \begin{alignat}{4}
 	&{{\mu_R}^e}_e=1,&\quad &{{\mu_R}^q}_e=0,&\quad &{{\mu_R}^p}_e=0,&\quad &{{\mu_R}^\alpha}_e=0,\\
 	&{{\mu_R}^e}_q=\frac{1}{2}(p\cos\alpha-q\sin\alpha),&\quad  &{{\mu_R}^q}_q=\cos\alpha, &\quad &{{\mu_R}^p}_q=\sin\alpha, &\quad &{{\mu_R}^\alpha}_q=0,\\
 	&{{\mu_R}^e}_p=-\frac{1}{2}(p\sin\alpha+q\cos\alpha),&\quad & {{\mu_R}^q}_p=-\sin\alpha, &\quad &{{\mu_R}^p}_p=\cos\alpha,&\quad & {{\mu_R}^\alpha}_p=0,	\\
 	&{{\mu_R}^e}_\alpha=0,&\quad &{{\mu_R}^q}_\alpha=0, & \quad &{{\mu_R}^p}_\alpha=0,&\quad & {{\mu_R}^\alpha}_\alpha=1.
 \end{alignat}
Then eqs.~\eqref{a58} take the form:
\begin{align}
	& \frac{\mathrm{d}e}{\mathrm{d}t}=\Pi^e+\frac{1}{2}(p\cos\alpha-q\sin\alpha)\Pi^q-\frac{1}{2}(p\sin\alpha+q\cos\alpha)\Pi^p,\\
	& \frac{\mathrm{d}q}{\mathrm{d}t}=\cos\alpha\,\Pi^q-\sin\alpha\,\Pi^p,\\
	& \frac{\mathrm{d}p}{\mathrm{d}t}=\sin\alpha\,\Pi^q+\cos\alpha\,\Pi^p,\\
& \frac{\mathrm{d}\alpha}{\mathrm{d}t}=\Pi^\alpha.	
\end{align}
Using \eqref{a63}$\div$\eqref{a66}, together with the initial conditions $e(0)=q(0)=p(0)=\alpha(0)=0$ we arrive at the following solution to the geodesic equations:
\begin{align}
	& \alpha =Bt,\label{a76}\\
	&q=\frac{D}{\tilde{\nu}}\naw{\cos(\tilde{\nu}t)-1}+\frac{F}{\tilde{\nu}}\sin(\tilde{\nu}t),\label{a77}\\
	& p=\frac{D}{\tilde{\nu}}\sin(\tilde{\nu}t)+\frac{F}{\tilde{\nu}}(1-\cos(\tilde{\nu}t)),\label{a78}\\
	& e=\naw{A-\frac{D^2+F^2}{2\tilde{\nu}}}t+\frac{D^2+F^2}{2\tilde{\nu}^2}\sin(\tilde{\nu}t),\label{a79}\\
	& \tilde{\nu}\equiv aA+(b+2)B.\label{a80}	
\end{align}
Due to the scaling properties of eqs.~\eqref{a76}$\div$\eqref{a80} one may assume that the endpoint corresponds to $t=1$. This allows us to find the constants $A$, $B$, $D$, $F$ in terms of coordinates of endpoint
\begin{align}
	& A=\frac{\tilde{\nu}-(b+2)\alpha}{a},\label{a81}\\
	& B=\alpha,\label{a82}\\
	& D=\frac{\tilde{\nu}}{2}\naw{\frac{p\sin\tilde{\nu}}{1-\cos\tilde{\nu}}-q},\label{a83}\\
	& F=\frac{\tilde{\nu}}{2}\naw{\frac{q\sin\tilde{\nu}}{1-\cos\tilde{\nu}}+p},\label{a84}
\end{align}
where $\tilde{\nu}$ is a solution of the transcendental equation:
\begin{equation}
	e=\frac{\tilde{\nu}-(b+2)\alpha}{a}+\naw{\frac{q^2+p^2}{4}}\naw{\frac{\sin\tilde{\nu}-\tilde{\nu}}{1-\cos\tilde{\nu}}}.\label{a85}
\end{equation}
The length of the geodesic starting at the unit element and terminating at $g$ is:
\begin{equation}
	l(g)=\int\limits_0^1\sqrt{\eta_{ij}\Pi^i\Pi^j}dt=\sqrt{\frac{(\tilde{\nu}-2\alpha)^2+(ad-b^2)\alpha^2}{a}+\frac{\tilde{\nu}^2(q^2+p^2)}{2(1-\cos\tilde{\nu})}}.\label{a86}
\end{equation}
We conclude that, within the approach based on the geometry of group (rather than considering the explicit operator realization), the problem of finding $l(g)$ reduces to solving the transcendental equation \eqref{a85} and then computing the corresponding value of $l(g)$ from eq.~\eqref{a86}. The reason that one can find the solution in closed form is that, at final instance, we are dealing with Hamiltonian system with considerably large symmetry. However, we encounter here two obstacles:
\begin{itemize}
	\item[(i)] eq.~\eqref{a85} cannot be, in general, solved analytically;
	\item[(ii)] for a given endpoint $g=(e,q,p,\alpha)$ eq.~\eqref{a85} admits usually more than one solution; for our purposes we are interested in the minimal value of $l(g)$. 
\end{itemize}
Actually, it appears that eq.~\eqref{a85} admits, for $g=(e,q,p,\alpha)$ running over the whole group manifold, rich variety of solutions. The number of solutions can be finite or infinite while the minimum of $l(g)$ can be attained for more or less ''trivial'' solutions of \eqref{a85}. This analysis, although it does not refer to specific physical systems, is important since it shows that, in general, Nielsen's complexity can be determined only taking into account global structure of symmetry group. One should also keep in mind that the choice of representation may be also important: for homomorphic representations the classification of solutions on $G/K$, $K$ being the kernel of representation, are relevant. Global properties cannot be uncovered if one refers only to approximate solutions.

In order to analyze \eqref{a85} in more detail let us define 

\begin{equation}
	\Delta\equiv\frac{4}{a(q^2+p^2)},\quad \Gamma\equiv\naw{e+\frac{(b+2)\alpha}{a}}\naw{\frac{4}{q^2+p^2}};\label{a87}
	\end{equation}
we have assumed here that $q^2+p^2\neq 0$ since in the case $q^2+p^2=0$ eq.~\eqref{a85} reduces significantly. As $g=(e,q,p,\alpha)$ varies over group manifold, $\Delta$ and $\Gamma$ are functionally independent. With the notation \eqref{a87}, eq.~\eqref{a85} takes the form:
\begin{equation}
	\Gamma=\Delta\tilde{\nu}+\frac{\sin\tilde{\nu}-\tilde{\nu}}{1-\cos\tilde{\nu}}.\label{a88}
\end{equation}
Note that $\Delta\in(0,\infty)$ while $\Gamma\in\mathbbm{R}$. 

Denoting by $f(\tilde{\nu};\Delta)$ the right-hand side of eq.~\eqref{a88} we see that $f(-\tilde{\nu};\Delta)=-f(\tilde{\nu};\Delta)$, $f(0;\Delta)=0$, $f(\tilde{\nu};\Delta)\underset{\tilde{\nu}\rightarrow 2\pi^-}\longrightarrow -\infty$, $f(\tilde{\nu};\Delta)\underset{\tilde{\nu}\rightarrow -2\pi^+}\longrightarrow\infty$.

\begin{figure}[h!]
	\begin{center}
		\subfigure[$0<\Delta<\frac{1}{3}$ ]{\label{fig1a}\includegraphics[width=0.45\textwidth]{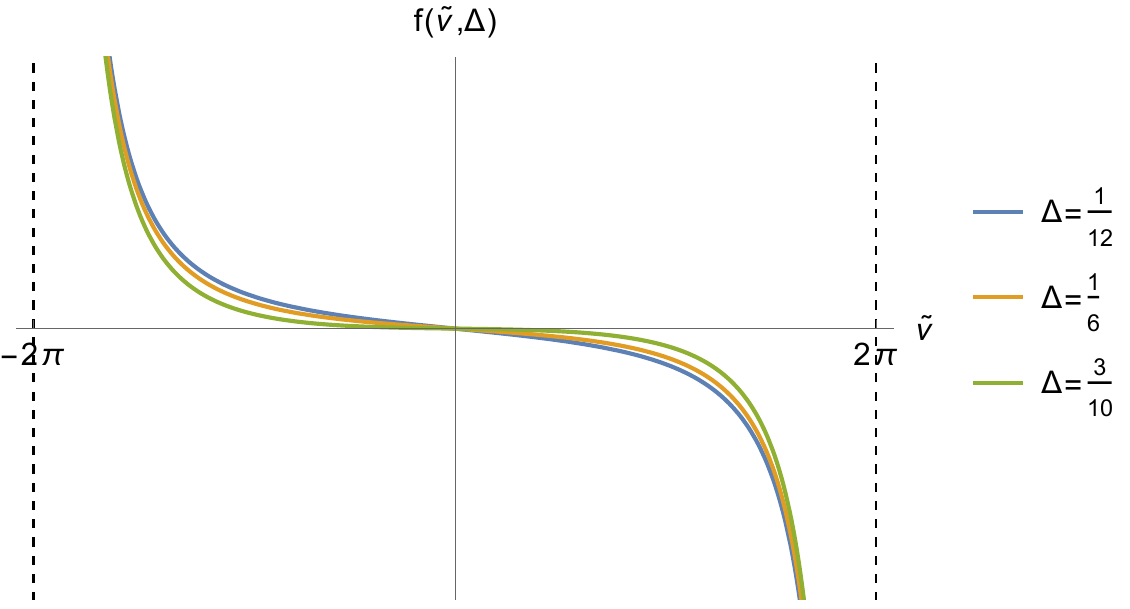}}\qquad
		\subfigure[$\Delta=\frac{1}{3}$]{\label{fig1b}\includegraphics[width=0.45\textwidth]{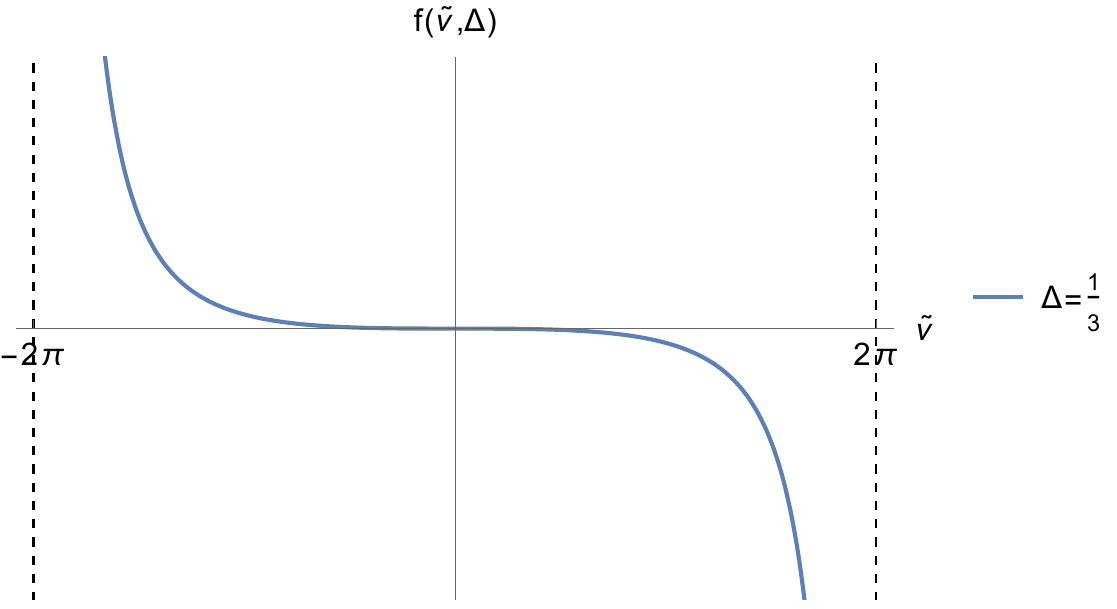}}\\
		\subfigure[$\Delta>\frac{1}{3}$]{\label{fig1c}\includegraphics[width=0.45\textwidth]{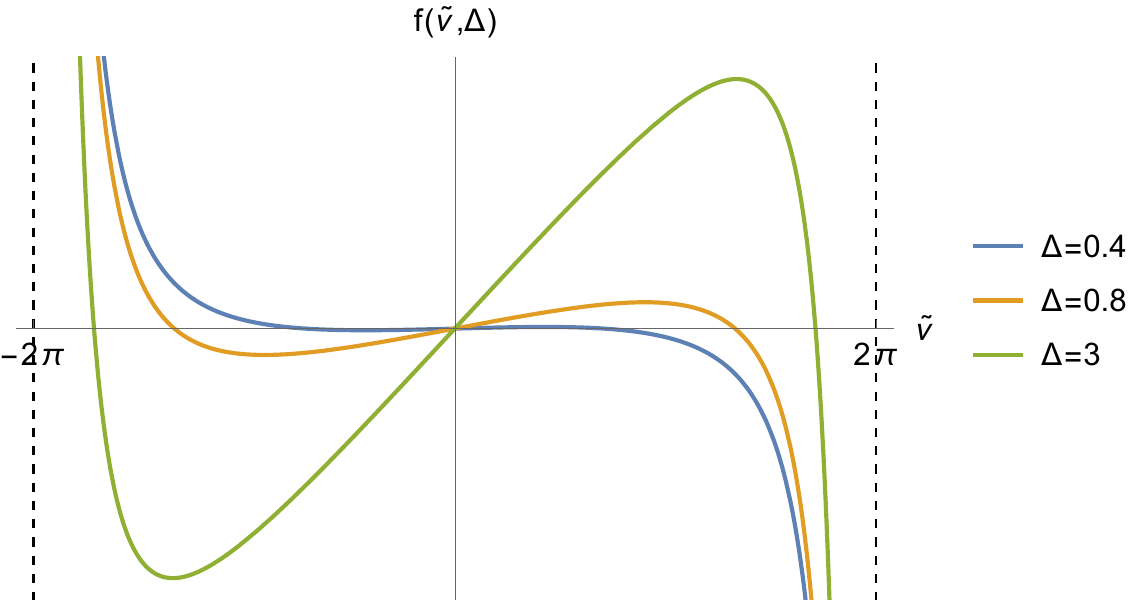}}
	\end{center}
	\caption{The shape of $f(\tilde{\nu};\Delta)$ for different value of $\Delta$.} 	
\end{figure}

In the interval $-2\pi <\tilde{\nu}<2\pi$ there is always one solution of eq.~\eqref{a85} for $0<\Delta\leq \frac{1}{3}$ (cf.~Fig.~\ref{fig1a} and \ref{fig1b}) and one, two or three solutions for $\Delta>\frac{1}{3}$ (Fig.~\ref{fig1c}), depending on the value of $\Gamma$. However, in general there are also solutions in the intervals $(2\pi k, 2\pi(k+1))$, $k=1,2,\dots$ or $k=-2,-3,\dots$. Once all solution are found one has to select the one corresponding to the minimal value of $l(g)$. It is not difficult to show that the minimal length $l(g)$ is not necessarily attained by the solution $\tilde{\nu}$ from the interval $(-2\pi,2\pi)$. Let us put $\tilde{\nu}=(2k+1)\pi$ with $k$ being an arbitrary integer. Then \eqref{a86} implies
\begin{equation}
	l^2(g)=\frac{1}{a}\naw{((2k+1)\pi-2\alpha)^2+(ad-b^2)\alpha^2+\frac{1}{\Delta}(2k+1)^2\pi^2}.\label{a89}
\end{equation} 
Let us fix an integer $k\geq 1$. Putting
\begin{equation}
	\alpha=(2k+1)\naw{\frac{\Delta+1}{\Delta}}\frac{\pi}{2},
\end{equation}
we see that $\tilde{\nu}=(2k+1)\pi$ provides the absolute minimum of
\begin{equation}
	{\tilde{l}}\,^2(g)=\frac{1}{a}\naw{(\tilde{\nu}-2\alpha)^2+\alpha^2(ad-b^2)+\frac{\tilde{\nu}^2}{\Delta}}.\label{a91}
\end{equation}
On the other hand
\begin{equation}
	l^2(g)\geq \tilde{l}\,^2(g),\label{a92}
\end{equation}
and the equality holds for $\tilde{\nu}=(2m+1)\pi$, with $m$ integer. Now, setting $\tilde{\nu}=(2k+1)\pi$ in eq.~\eqref{a88} we find the relevant $\Gamma$, i.e.~the value of $e$. So we have found an element $g$ for which the minimum of $l(g)$ attained for $\tilde{\nu}=(2k+1)\pi$ with arbitrary $k\geq 1$.

\begin{figure}[!h]
	\centering
	\includegraphics[scale=0.6]{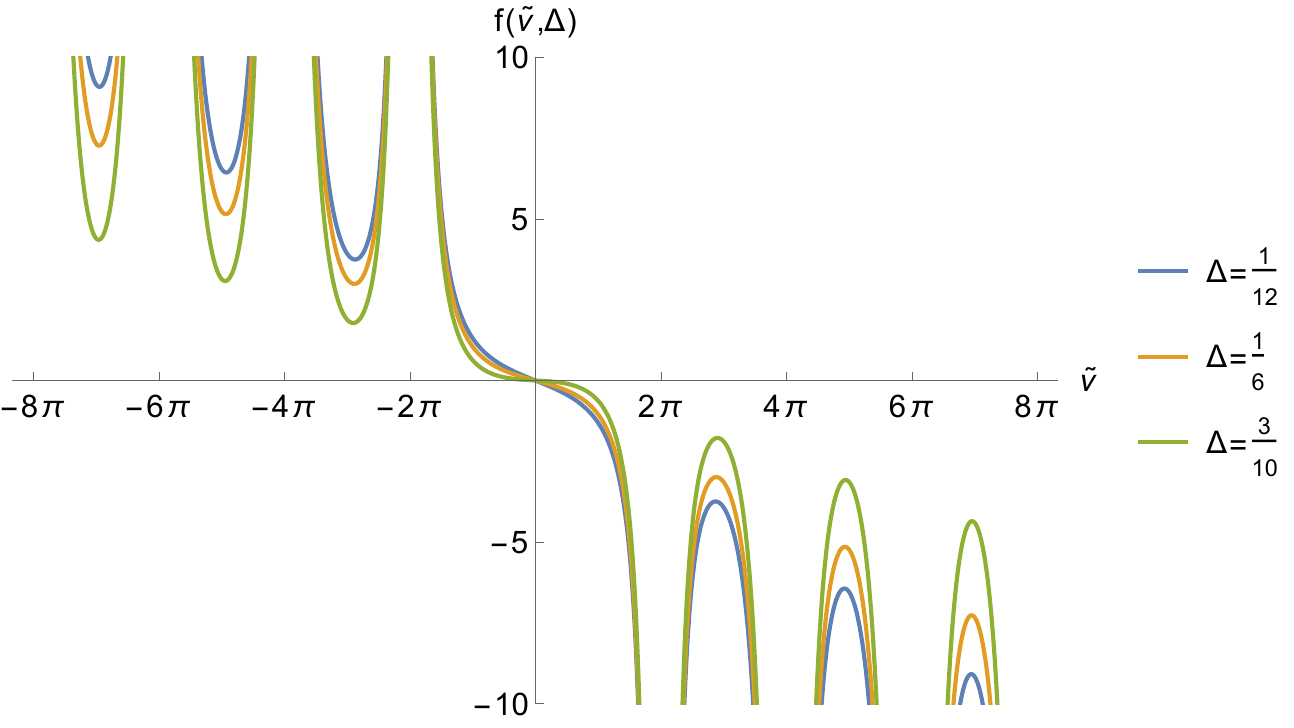}
	\caption{The shape of $f(\tilde{\nu},\Delta)$ for $\Delta<\frac{1}{2}$.}\label{fig2}
\end{figure}

\begin{figure}[!h]
	\centering
	\includegraphics[scale=0.6]{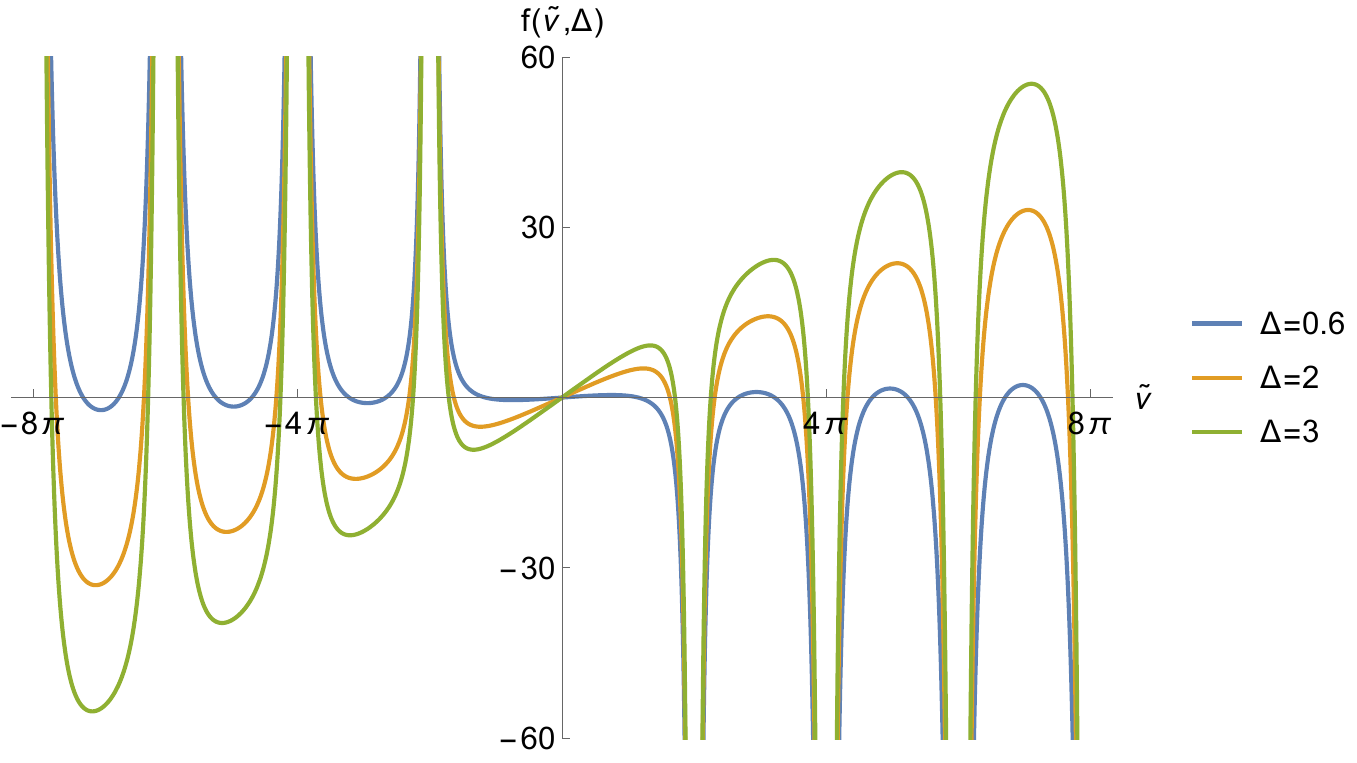}
	\caption{The shape of $f(\tilde{\nu},\Delta)$ for $\Delta>\frac{1}{2}$.}\label{fig3}
\end{figure}

To gain some information about the number of geodesics connecting $e$ with a given element $g$ consider eq.~\eqref{a88} in more detail. The function $f(\tilde{\nu};\Delta )$ is antisymmetric, $f(-\tilde{\nu};\Delta)=-f(\tilde{\nu};\Delta )$, so it is sufficient to consider the case of positive $\tilde{\nu}$. The interval $\left<0,2\pi\right)$ was already described above. Consider now the interval $(2\pi k,2\pi(k+1))$, $k=1,2,\dots$; $f(\tilde{\nu};\Delta )$ has two vertical asymptotics, $\tilde{\nu}=2\pi k$, $\tilde{\nu}=2\pi (k+1)$ and  $f(\tilde{\nu};\Delta )\rightarrow -\infty$ for $\tilde{\nu}\rightarrow 2\pi k^+$ and $\tilde{\nu}\rightarrow 2\pi(k+1)^-$. Moreover, it is concave in any interval $(2\pi k,2\pi(k+1))$ so it has exactly one maximum. It corresponds to $\tilde{\nu}$ obeying  
\begin{equation}
	\tilde{\nu}=\frac{2(1-\cos\tilde{\nu})-\Delta(1-\cos\tilde{\nu})^2}{\sin{\tilde{\nu}}}. \label{b1}
\end{equation}
Let us derive the asymptotic formula for $\tilde{\nu}$ when $k\rightarrow\infty$. Denote by $\tilde{\nu}_k$ the solution to eq.~\eqref{b1}
in the interval $(2\pi k,2\pi(k+1))$. For large $\tilde{\nu}$ eq.~\eqref{b1} tells us that $\tilde{\nu}_k$ must be close to $(2k+1)\pi$. We put 
\begin{equation}
	\tilde{\nu}_k=(2k+1)\pi-\delta_k. \label{b2}
\end{equation}
Then \eqref{b1} takes the approximate form 
\begin{equation}
	(2k+1)\pi-\delta_k \cong\frac{4(1-\Delta)+(2\Delta-1)\delta^2_k}{\delta_k}, \label{b3}
\end{equation}
which yields
\begin{equation}
	\tilde{\nu}_k=(2k+1)\pi-\frac{4(1-\Delta)}{(2k+1)\pi}+O\naw{\frac{1}{k^2}},\label{b4}
\end{equation}
and
\begin{equation}
	f(\tilde{\nu};\Delta )=\naw{\Delta-\frac{1}{2}}(2k+1)\pi+\frac{2(1-\Delta)(2-\Delta)}{(2k+1)\pi}+O\naw{\frac{1}{k^2}}.\label{b5}
\end{equation}
	Therefore, the maxima of $f(\tilde{\nu};\Delta )$ lie, for $\tilde{\nu}>0$, asymptotically on the straight line
	\begin{equation}
		f(\tilde{\nu};\Delta )=(\Delta-\frac{1}{2})\tilde{\nu}.\label{b98}
	\end{equation}
Taking into account that $f(\tilde{\nu};\Delta )$ is odd we conclude that the eq.~\eqref{a88} has a finite number of solutions for $0<\Delta<\frac{1}{2}$ and infinite for $\Delta>\frac{1}{2}$, cf.~Figs. \ref{fig2} and \ref{fig3}.


	The above brief analysis shows that the pattern of geodesics on oscillator group is quite complicated. Given a right-invariant metric, characterized by the parameters $a$, $b$, $d$ (cf.~eq.~\eqref{a54}) and an endpoint $g=(e,q,p,\alpha)$, the geodesics connecting $g$ with the unit element are determined by the solutions of eq.~\eqref{a88} (i.e., basically, eq.~\eqref{a85}); the corresponding lengths $l(g)$ are given by \eqref{a86}. Two main conclusions read:
	\begin{itemize}
		\item[-] the number of solutions to eq.~\eqref{a85} (i.e.~the number of geodesics) is finite for $0<\Delta<\frac{1}{2}$ and infinite for $\Delta>\frac{1}{2}$; moreover, for $0<\Delta\leq\frac{1}{3}$ there exists unique solution in the interval $-2\pi<\tilde{\nu}<2\pi$ while for $\Delta>\frac{1}{3}$ there are one, two or three solutions in this interval, depending on the value of $\Gamma$;
		\item[-] for specific endpoints $g$ the minimal values of $l(g)$ are attained in the intervals $(2\pi k, 2\pi(k+1))$, with $k$ being arbitrary integer.  
	\end{itemize}

\subsection{Interpretation in terms of Hamiltonian dynamics}	 

It is nice to understand the dynamics of geodesics from a slightly different point of view. For a general right-invariant metric (general $\eta$) there exist four integrals of motion related, via Noether theorem, to the right $G$-action on itself. However, they obey (with respect to the Poisson bracket) the commutation rules of the Lie algebra of $G$ (see Appendix). Therefore, the complete integrability of equations describing the geodesics cannot be taken for granted; the Arnold-Liouville integrability requires the existence of integrals in involution - the integrability conditions in the non-abelian case are more complicated \cite{Mishchenko}.

For general $\eta$ one can choose two commuting integrals (say $E$ and $H$) resulting from right action of $G$ on itself. Additionally, ``energy'' is conserved since the geodesic Lagrangian does not depend explicitly on the affine evolution parameter. Therefore, one has, in general, three (commuting) integrals of motion and still one, commuting with them, is lacking to guarantee integrability. It can be found by noting that the metric defined by $\eta$ given by eq.~\eqref{a54} enjoys symmetry under the rotations in $q-p$ plane. As a result, we get the fourth integral, ``angular momentum'', making our dynamics completely integrable. 

The right-invariant length element for the metric \eqref{a54} takes the form (cf.~eq.~\eqref{Ap27}):
\begin{equation}
	\mathrm{d}s^2 =a\mathrm{d}e^2+\mathrm{d}q^2+\mathrm{d}p^2+\frac{a}{4}\naw{q\mathrm{d}p-p\mathrm{d}q}^2+2b\mathrm{d}e\mathrm{d}\alpha+d\mathrm{d}\alpha^2+(q\mathrm{d}p-p\mathrm{d}q)(a\mathrm{d}e+b\mathrm{d}\alpha).\label{a93}
\end{equation}
The relevant Lagrangian reads:
\begin{equation}
	L=\frac{1}{2}\naw{a\dot{e}^2+\dot{q}^2+\dot{p}^2+\frac{a}{4}(q\dot{p}-p\dot{q})^2+2b\dot{e}\dot{\alpha}+d\dot{\alpha}^2+(q\dot{p}-p\dot{q})(a\dot{e}+b\dot{\alpha})}.\label{a94}
\end{equation}
We see that $e$ and $\alpha$ are cyclic variables so the corresponding momenta are conserved:
\begin{align}
	& p_e\equiv\frac{\partial L}{\partial\dot{e}}=a\dot{e}+b\dot{\alpha}+\frac{a}{2}(q\dot{p}-p\dot{q}),\label{a95}\\
		& p_\alpha\equiv\frac{\partial L}{\partial\dot{\alpha}}=b\dot{e}+d\dot{\alpha}+\frac{b}{2}(q\dot{p}-p\dot{q}).\label{a96}
\end{align}
Moreover, the rotations in $q-p$ plane, $\delta q=p\delta\lambda$, $\delta p=-q\delta\lambda$ are symmetries and yield the integral of motion
\begin{equation}
	J=-\frac{1}{2}(q^2+p^2)(a\dot{e}+b\dot{\alpha})+\naw{1+\frac{a}{4}(q^2+p^2)}(p\dot{q}-q\dot{p}).\label{a97}
\end{equation}
Therefore, we have four commuting integrals of motion, $E=L$, $p_e$, $p_\alpha$ and $J$; our dynamics integrable in the Arnold-Liouville sense. In order to reproduce the solutions \eqref{a76}$\div$\eqref{a79} we first solve eqs.~\eqref{a95}, \eqref{a96} for $\dot{e}$ and $\dot{\alpha}$:
\begin{align}
&	\dot{e}=\frac{1}{2}(p\dot{q}-q\dot{p})+\frac{dp_e-bp_\alpha}{ad-b^2},\label{a98}\\
&	\dot{\alpha}=\frac{ap_\alpha-bp_e}{ad-b^2}.\label{a99}
\end{align}  
Inserting \eqref{a98} and \eqref{a99} into \eqref{a97} one finds
\begin{equation}
	J=p\dot{q}-q\dot{p}-\frac{p_e}{2}(q^2+p^2).\label{a100}
\end{equation}
Let us introduce the polar coordinates in $q-p$ plane:
\begin{align}
	&q=r\cos\gamma,\label{a101}\\
	&p=r\sin\gamma.\label{a102}
\end{align}
Then
\begin{equation}
	J=-r^2\dot{\gamma}-\frac{p_e}{2}r^2,\label{a103}
\end{equation}
while the ``energy'' reads:
\begin{equation}
	\mathcal{E}\equiv L=\frac{\dot{r}^2}{2}+\frac{J^2}{2r^2}+\frac{p_e^2}{8}r^2+\frac{Jp_e}{2}+\frac{(dp_e-bp_\alpha)p_e+(ap_\alpha-bp_e)p_\alpha}{2(ad-b^2)}.\label{a104}
\end{equation}
Therefore, as far as $q$ and $p$ are concerned, we are dealing with two-dimensional harmonic oscillator with the frequency $\frac{\modu{p_e}}{2}$.\footnote{See, however, the remark below.} Once eq.~\eqref{a104} is solved by separating variables one can find $\gamma$ from eq.~\eqref{a103} and, finally, $e$ and $\alpha$ from eqs.~\eqref{a98} and \eqref{a99}.
Due to the initial conditions $q(0)=0=p(0)$ one concludes that $J=0$ and our equations simplify to:
\begin{align}
	& \dot{e}=\frac{p_e}{4}r^2+\frac{dp_e-bp_\alpha}{ad-b^2},\label{a105}\\
	&\dot{\alpha}=\frac{ap_\alpha-bp_e}{ad-b^2},\label{a106}\\
	& \dot{\gamma}=-\frac{p_e}{2},\label{a107}\\
	&\dot{r}^2=2\mathcal{E}-\naw{\frac{p_e}{2}}^2r^2-\frac{p_e(dp_e-bp_\alpha)+p_\alpha(ap_\alpha-bp_e)}{ad-b^2}.\label{a108}
\end{align}
The solution with the initial conditions $r(0)=0$, $e(0)=0$, $\alpha(0)=0$ and $\gamma(0)=\gamma_0$ takes the form:
\begin{align}
	& q=\frac{\mathcal{A}}{2}\cos\gamma_0\sin(p_e\sigma)+\frac{\mathcal{A}}{2}\sin\gamma_0(1-\cos(p_e\sigma)),\label{a109}\\
&	p=-\frac{\mathcal{A}}{2}\cos\gamma_0(1-\cos(p_e\sigma))+\frac{\mathcal{A}}{2}\sin\gamma_0\sin(p_e\sigma),\label{a110}\\
& \alpha=\naw{\frac{ap_\alpha-bp_e}{ad-b^2}}\sigma,\label{a111}\\
& e=-\frac{\mathcal{A}^2}{8}\sin(p_e\sigma)+\naw{\frac{p_e\mathcal{A}^2}{8}+\frac{dp_e-bp_\alpha}{ad-b^2}}\sigma,\label{a112}
\end{align}
with
\begin{equation}
	\mathcal{A}^2\equiv\frac{8\mathcal{E}}{p_e^2}-4\naw{\frac{p_e(dp_e-bp_\alpha)+p_\alpha(ap_\alpha-bp_e)}{p_e^2(ad-b^2)}}.\label{a113}
\end{equation}
This can be compared with eqs.~\eqref{a76}$\div$\eqref{a79}. The latter coincide with eqs.~\eqref{a109}$\div$\eqref{a112} provided $F/\tilde{\nu}=-\frac{\mathcal{A}}{2}\cos\gamma_0$, $D/\tilde{\nu}=-\frac{\mathcal{A}}{2}\sin\gamma_0$, $\tilde{\nu}=-p_e$, $B=\frac{ap_\alpha-bp_e}{ad-b^2}$. The advantage of the above alternative approach is that the length of the relevant geodesics is given simply by $\sqrt{2\mathcal{E}}$. Moreover, by performing complete Hamiltonian analysis one finds that the $q-p$ dynamics may be also viewed as the plane motion of charged particle in uniform magnetic field proportional to $p_e$. In fact, $H$ takes the form
\begin{equation}
	H=\frac{1}{2}\com{\naw{p_p-\frac{1}{2}qp_e}^2+\naw{p_q+\frac{1}{2}pp_e}^2}+\frac{ap_\alpha^2+dp_e^2-2bp_\alpha p_e}{2(ad-b^2)},
\end{equation} 
and is related to $C_2$ by the formula:
\begin{equation}
	C_2=-H+p_\alpha p_e+\frac{ap_\alpha^2+dp_e^2-2bp_\alpha p_e}{ad-b^2}.
\end{equation}

\section{Complexity}\label{sec4}

Basing on the results obtained above we analyse now the Nielsen complexity $\mathcal{C}$ of some observables (according to Nielsen's et al.~results $\mathcal{C}$ provides a lower bound on gate complexity). To this end let us select an unitary representation of the oscillator group corresponding to the following values of Casimir operators:
\begin{align}
	&C_1\equiv E=\Omega I, \quad \Omega>0, \label{a114}\\
	&C_2\equiv HE-\frac{1}{2}(Q^2+P^2)=hI.\label{a115}
\end{align}
Defining:
\begin{align}
& \widetilde{Q}\equiv\frac{Q}{\Omega},\label{a116}\\
&\widetilde{P}=P,\label{a117}
\end{align}
one finds
\begin{align}
&	\com{\widetilde{Q},\widetilde{P}}=iI,\label{a118}\\
& H=\frac{1}{2\Omega}(\widetilde{P}^2+\Omega^2\widetilde{Q}^2)+\frac{h}{\Omega}I.\label{a119}
\end{align}
Therefore $\widetilde{Q}$ and $\widetilde{P}$ are canonically conjugated variables while
\begin{equation}
	H_{osc}\equiv\Omega H=\frac{1}{2}(\widetilde{P}^2+\Omega^2\widetilde{Q}^2)+hI, \label{a120}
\end{equation}
represents harmonic oscillator of the frequency $\Omega$ and ground state energy shifted by $h$. 

The evolution operator
\begin{equation}
	e^{-itH_{osc}}=e^{-it\Omega H}, \label{a121}
\end{equation}
represents the element, within the unitary representation under consideration, with $e=q=p=0$ and $\alpha=-\Omega t$. Then we find from eq.~\eqref{a85}
\begin{equation}
	\tilde{\nu}=(b+2)\alpha=-(b+2)\Omega t, \label{a122}
\end{equation}
while \eqref{a81}$\div$\eqref{a84} yield
\begin{equation}
	A=0,\quad D=F=0,\quad B=-\Omega t.
\end{equation}
Then the solution to the geodesic equations \eqref{a76}$\div$\eqref{a79} takes the form
\begin{align}
	& \alpha(\sigma)=-\Omega t\sigma, \label{a124}\\
	& q(\sigma)=p(\sigma)=e(\sigma)=0.\label{a125}
\end{align}
Therefore, the length of the geodesics equals, according to the equation \eqref{a86}:
\begin{equation}
	l(g)=\sqrt{d}\modu{\Omega t}.\label{a126}
\end{equation}
However, one should keep in mind that we are dealing with the faithful representation of $\tilde{G}/K$ with $K$ being the kernel of the representation. Now, the spectrum of $E$ consists of single point $\Omega$, the spectra of $Q$ and $P$ extend over the whole axis while that of $H$ reads
\begin{equation}
	h_n=n+\frac{1}{2}+\frac{h}{\Omega},\quad n=1,2,\ldots \label{a127}
\end{equation}
For $\frac{h}{\Omega}$ irrational the kernel consists of the elements $\naw{\exp\naw{\frac{2\pi i}{\Omega}E}}^k$ with integer $k$. The topology of $G/K$ is $S_e\times\mathbbm{R}_\alpha\times\mathbbm{R_q}\times\mathbbm{R_p}$ and the complexity is given by eq.~\eqref{a126}. On the other hand, if $\frac{h}{\Omega}+\frac{1}{2}=\frac{k}{l}$, $K$ consists of the elements
\begin{equation}
	g_{km}=\naw{\exp\naw{\frac{2\pi i E}{\Omega}}}^k\naw{\exp\naw{2\pi ilH}}^m, \label{a128}
\end{equation} 
and the topology of $G/K$ is $S_e\times S_\alpha\times \mathbbm{R}_q\times \mathbbm{R}_p$. The topology must then be taken into account when computing the complexity. In particular, for $h=0$ one finds
\begin{equation}
	\mathcal{C}=\sqrt{d}\begin{cases}
		\Omega(t-4\pi k) & \text{for} \quad 4\pi k\leq t<4\pi k+2\pi,\\
			\Omega(4\pi (k+1)-t) & \text{for} \quad 4\pi k+2\pi\leq t<4\pi (k+1),
	\end{cases}\label{a129}
\end{equation}
in agreement with \cite{Chowdhury}.

As the second example consider \cite{Chowdhury}
\begin{equation}
U(g(0,0,q,p))=\exp(i(pQ+qP)). \label{a130}
\end{equation}
Eqs.~\eqref{a85}, \eqref{a86} take the form
\begin{align}
&	\frac{\tilde{\nu}}{a}+\naw{\frac{q^2+p^2}{4}}\naw{\frac{\sin\tilde{\nu}-\tilde{\nu}}{1-\cos\tilde{\nu}}}=0, \label{a131}\\
& l(g)=\sqrt{\frac{\tilde{\nu}^2}{a}+\naw{\frac{q^2+p^2}{2}}\frac{\tilde{\nu}^2}{1-\cos\tilde{\nu}}}, \label{a132}	
\end{align}
respectively. Eqs.~\eqref{a131} admits the solution $\tilde{\nu}=0$ leading to
\begin{equation}
	l(g)=\sqrt{q^2+p^2}.\label{a133}
\end{equation}
This is the minimal value of geodesic length. Indeed, the second term under the square root on the right hand side of eq.~\eqref{a132} attains its minimum at $\tilde{\nu}=0$. Therefore
\begin{equation}
	\mathcal{C}=\sqrt{q^2+p^2},\label{a134}
\end{equation}
which again agrees with the result of \cite{Chowdhury}. 

Consider now the harmonic oscillator perturbed by linear therm:
\begin{align}
&	U(t)=\exp(-it\widetilde{H}), \label{a135}\\
	& \widetilde{H}=\frac{1}{2}(\widetilde{P}^2+\Omega^2\widetilde{Q}^2)+\lambda\widetilde{Q}.\label{a136}
\end{align}
Choosing again the representation of oscillator group with $E=\Omega I$, $h=0$ and redefining $Q\equiv\Omega\widetilde{Q}$, $P\equiv\widetilde{P}$ one can put $U(t)$ into the form
\begin{equation}
	U(t)=\exp\naw{-i\naw{\Omega H+\frac{\lambda Q}{\Omega}}}.\label{a137}
\end{equation}
In order to compute the complexity of $U(t)$ we have first to factorize $\exp(i(\alpha H+pQ))$. This can be easily done by writing out a simple differential equation and using Magnus formula. Alternatively, one can use the results concerning exponential mapping (eqs.~\eqref{a20}$\div$\eqref{a27}).

The result reads:
\begin{equation}
	\begin{split}
	U(t)=&\exp\naw{\frac{i\lambda^2}{2\Omega^4}\naw{\Omega t-\sin(\Omega t)}E}\exp\naw{-it\Omega H}\\
	&\cdot\exp\naw{\frac{-i\lambda}{\Omega^2}\sin(\Omega t)Q+\frac{i\lambda}{\Omega^2}(\cos(\Omega t)-1)P}.\label{a138}
\end{split}\end{equation}
Eq.~\eqref{a85} takes the form
\begin{equation}
\frac{\lambda^2}{2\Omega^4}\naw{\Omega t-\sin(\Omega t)}=\frac{\tilde{\nu}+(b+2)\Omega t}{a}+\frac{\lambda^2}{2\Omega^4}\frac{(1-\cos(\Omega t))(\sin\tilde{\nu}-\tilde{\nu})}{1-\cos\tilde{\nu}},\label{a139}
\end{equation}
while
\begin{equation}
	l(g)=\sqrt{\frac{(\tilde{\nu}+2\Omega t)^2+(ad-b^2)(\Omega t)^2}{a}+\frac{\lambda^2}{\Omega^4}\frac{\tilde{\nu}^2(1-\cos(\Omega t))}{(1-\cos\tilde{\nu})}}.\label{a140}
\end{equation}
Eq.~\eqref{a139} is, in general, a transcendental equation, difficult to solve. However, for the particular value $b=-1$ there exists a simple solution
\begin{equation}
	\tilde{\nu}=-\Omega t.\label{a141}
\end{equation}
Then
\begin{equation}
	l(g)=\modu{\Omega t}\sqrt{d+\frac{\lambda^2}{\Omega^4}}, \label{a142}
\end{equation}
which can be compared with the estimate presented in \cite{Chowdhury}.

	It is crucial to realize that eq.~\eqref{a142} does not necessarily give the value of complexity if eq.~\eqref{a149} admits more solutions. This can be shown explicitly: it appears that for some values of $\frac{\lambda}{\Omega^2}$ and $\Omega t$ these additional solutions (not accessible analytically, only numerically) yield smaller values of $l(g)$. We quote below in some detail numerical evidence supporting this conclusion in order to stress important fact that in order to determine correct value of complexity one has to take into account the global structure of symmetry group.

	 Let us put $a=1$, $d=2$ (keeping in mind that $b=-1$). 
The values of $\Delta$ and $\Gamma$, entering eq.~\eqref{a88}	read
\begin{align}
	& \Delta=\frac{2}{\frac{\lambda^2}{\Omega^4}(1-\cos(\Omega t))},\label{b148}\\
	& \Gamma=\frac{2\naw{\naw{\frac{\lambda^2}{2\Omega^4}-1}\Omega t}-\frac{\lambda^2}{2\Omega^4}\sin(\Omega t)}{\frac{\lambda^2}{\Omega^4}\naw{1-\cos(\Omega t)}}.\label{b149}
\end{align}
We have computed $l(g)$, eq.~\eqref{a140}, for few values of the parameters $\frac{\lambda^2}{\Omega^4}$ and $\Omega t$. We summarize below the results obtained.
\begin{itemize}
	\item[-] For $\frac{\lambda^2}{\Omega^4}=50$, $\Omega t=1$ there exists exactly one solution (this agrees with the value of $\Delta=0.0870<\frac{1}{2}$), $\tilde{\nu}=-1$, given by eq.~\eqref{a141}. Then eq.~\eqref{a140} gives the actual value of complexity, $\mathcal{C}\equiv l(g)=7.2111$.
	
	\item[-] For $\frac{\lambda^2}{\Omega^4}=10$, $\Omega t=1$ ($\Delta=0.4351<\frac{1}{2}$) there are three solutions:
	$\tilde{\nu}=-1$, $l(g)=3.464$; $\tilde{\nu}=-2.116$, $l(g)=3.817$; $\tilde{\nu}=2.905$, $l(g)=6.688$.
	Therefore, again the minimal value of $l(g)$ is attained for $\tilde{\nu}$ given by eq.~\eqref{a141} yielding the complexity $\mathcal{C}=3.464$.
	
	\item[-] For $\frac{\lambda^2}{\Omega^4}=10$, $\Omega t=10$ ($\Delta=0.1088<\frac{1}{2}$) we have again three solutions: 	$\tilde{\nu}=-10$, $l(g)=34.641$; $\tilde{\nu}=-8.162$, $l(g)=34.359$; $\tilde{\nu}=-4.621$, $l(g)=26.391$. Now the minimal value of $l(g)$ is attained for $\tilde{\nu}=-4.621$ leading to $\mathcal{C}=26.391$.
	
	\item[-] For $\frac{\lambda^2}{\Omega^4}=50$, $\Omega t=10$ ($\Delta=0.0218<\frac{1}{2}$) there exist three solutions: $\tilde{\nu}=-10$, $l(g)=72.111$; $\tilde{\nu}=-8.112$, $l(g)=71.148$; $\tilde{\nu}=-4.698$, $l(g)=48.325$. The complexity equals $\mathcal{C}=48.325$ and is attained for $\tilde{\nu}=-4.698$.
	
\item[-] For $\frac{\lambda^2}{\Omega^4}=50$, $\Omega t=50$ ($\Delta=1.1418>\frac{1}{2}$) there exists the infinite set of solutions; $\tilde{\nu}=-50$ corresponds to $l(g)=360.555$. However, there exists the solution $\tilde{\nu}=-6.188$ yielding $l(g)=161.500$; the latter is likely the minimal.

\item[-] For $\frac{\lambda^2}{\Omega^4}=10$, $\Omega t=50$ ($\Delta=5.7087>\frac{1}{2}$) again we are dealing with the infinite number of solutions;  $\tilde{\nu}=-50$ gives $l(g)=173.205$ while  $\tilde{\nu}=-6.180$ corresponds to $l(g)=117.579$ which is likely the correct value of complexity. 	
\end{itemize}

It would be nice to understand why $b=-1$ is a special case such that eq.~\eqref{a139} admits an explicit solution. Using \eqref{a76}$\div$\eqref{a84} we find immediately that for $b=-1$, $\tilde{\nu}=-\Omega t$ the geodesic is given by exponential mapping. This can be also verified from the general condition for the geodesic to be given by exponential mapping, derived in Appendix. To this end let us put
\begin{equation}
	X=eE+pQ+qP+\alpha H.\label{a143}
\end{equation}
Using eq.~\eqref{Ap55} we conclude that $\exp(i\sigma X)$ defines a geodesic curve provided
\begin{align}
&	q(ae+(b+1)\alpha)=0,\label{a144}\\
& p(ae+(b+1)\alpha)=0.\label{a145}
\end{align}
Therefore, either $q^2+p^2=0$ or $ae+(b+1)\alpha=0$. Eq.~\eqref{a121} corresponds to $q=0=p$ while eq.~\eqref{a130} - to $e=0=\alpha$. In the case of shifted oscillator, eq.~\eqref{a137}, $q\neq 0$, $e=0$; therefore, eq.~\eqref{a144} implies $b=-1$.

Assume now that $b\neq -1$ but $U(t)$, entering eq.~\eqref{a137}, contains an additional term $\exp(ieE)$. Then eq.~\eqref{a139} is replaced by
\begin{equation}
e+\frac{\lambda^2}{2\Omega^4}\naw{\Omega t-\sin(\Omega t)}=\frac{\tilde{\nu}+(b+2)\Omega t}{a}+\frac{\lambda^2}{2\Omega^4}\frac{(1-\cos(\Omega t))(\sin\tilde{\nu}-\tilde{\nu})}{1-\cos\tilde{\nu}}.\label{a146}
\end{equation}
It admits again the solution $\tilde{\nu}=-\Omega t$ provided $ae+(b+1)\alpha=0$.

Let us note that the relations \eqref{a144}, \eqref{a145} are invariant under the authomorphisms \eqref{a28}$\div$\eqref{a31} preserving the form of the metric $\eta$ given by \eqref{a54}. Imposing the latter condition we find $\mu^2+\nu^2=1$, $\rho=\sigma=0$ so that
\begin{align}
	& e=e'+\tau\alpha',\label{a147}\\
	& p=\mu p'-\nu q',\label{a148}\\
	& q=\nu p'+\mu q',\label{a149}\\
	& \alpha=\alpha', \label{a150}
\end{align}
while the metric transforms as:
\begin{align}
& 	a=a',\label{a151}\\
	& d=d'+a'\tau^2-2b'\tau,\label{a152}\\
	& b=b'-a'\tau.\label{a153}
\end{align}
It is easy to check that \eqref{a144}, \eqref{a145} are invariant under these transformations.

\section{Conclusions and outlook}\label{sec5}

For the $n$ qubit case the Hilbert space of states is finite-dimensional. The set of all unitaries (of unit determinant) form the defining representation of the $SU(2^n)$ group. This makes life both easier - one can work with a specific implementation rather than with abstract structure and harder - it becomes slightly more difficult to recognize the result following directly from the abstract framework. In fact, a closer look on basic results of Nielsen's approach reveals that the latter can be described in abstract terms. The universal gates correspond to the subset of the group $SU(2^n)$ such that any element of the group can be represented in form of the group product of finite number of elements belonging to this subset (represented exactly or approximately in the sense of the group topology). Nielsen's et al.~ results \cite{NielsenChuang,Nielsen,NielsenDowling} concerning the discrete and continuous approaches to complexity can be stated and proven at the abstract level referring only to the structures entering the definition of Lie group (alternatively, characterizing the particular $SU(2^n)$ group). The whole analysis can then be lifted to the level of Hilbert space and unitaries with the help of unitary representation just because the representation, by definition, reflects the algebraic and topological (if it is continuous) properties of the group. For the defining representation of a group $(SU(2^n)$ in our case) these two aspects are difficult to separate.

The infinite-dimensional case is more intriguing. If we want to cover the most general case we enter the realm of infinite-dimensional geometry. However, in general not all unitaries and the observables generating them are equally interesting. Those really important are related to the symmetries enjoyed by many physical systems. Even if the symmetry is broken if happens quite often that the Hamiltonian of the system have definite transformation properties under the action of some group; this is the case for spectrum generating algebras/groups. In any case, the relevant observables are related to the symmetry generators and have straightforward physical interpretation. We can assume that both the operators whose complexity we want to compute and the universal gates belong to the representation of symmetry group and the problem becomes purely group-theoretical. Once it is solved it allows to compute the complexity of a number of unitaries by choosing the appropriate unitary representations.

We applied the above ideas to the simple case of oscillator group. For the quite natural choice of right-invariant metric the geodesic equations define completely integrable dynamical system in the Arnold-Liouville sense. Therefore, they can be solved explicitly providing the quantitative expressions for complexity. The geodesic equations, when lifted to the operator level with the help of unitary representation, acquire the form which suggests the use of standard tools for solving operator evolution equations like, for example, the Dyson expansion. However, they can be solved on the level of Lie group manifold by applying the methods of Hamiltonian mechanics.

In order to illuminate some subtleties of the Nielsen approach let us go back to the evolution operator \eqref{a121}. Choosing a particular representation of the oscillator group we select definite values of the Casimir operators $C_1$ and $C_2$, eqs.~\eqref{a114} and \eqref{a115}, respectively. In particular, fixing $h$ we obtain the harmonic oscillator with the ground state energy $\frac{\Omega}{2}+h$. However, the same operator can be described by choosing another, inequivalent unitary irreducible representation, corresponding to $C_2=h'I$, $h'\neq h$ and adding to $H$ the element $E$  with appropriate coefficient. These different points of view on the same operator result, in general, in different complexities. This is because the complexities are defined with respect to the set of universal gates corresponding to definite group elements. Therefore, they are sensitive to the relative position, in group manifold, of the elements representing the gates and the one we are interested in.

	At the risk of repeating statements already made above let us recapitulate what we consider to be most important aspects of ``continuous'' approach to complexity.
We believe that the correct approach to the Nielsen complexity results from the following principles:
\begin{itemize}
	\item[-] among numerous candidates for observables only relatively few are physically relevant; they result via Noether - like algorithm from the assumed symmetries, exact, broken, spontaneously broken or gauge ones. These observables (or rather, unitaries obtained by exponentiating them) play the double role: we are interested in computing their complexity and some of them define universal gates;
	\item[-] Nielsen's complexity can be described entirely in terms of the geometry of symmetry group. Reference to the specific quantum system emerges on the level of particular representation of symmetry group. This implies, among other things, that if the same operator represents different group elements in different representations, the corresponding complexities ascribed to it are, in general, different;
	\item[-] the actual value of complexity depend on the global structure of the symmetry group and cannot be grasped if we refer only to approximate solutions to geodesic equations;
	\item[-] within the approach based on the geometry of group manifold we have at our disposal the tools from the theory of Hamiltonian systems with symmetries; this allows us to analyse effectively the relevant geodesic equations and exhibit the global properties of geodesics (for wider class of metrics). The analysis presented here can be quite straightforwardly extended to other groups, for example popular $SL(2,\mathbb{R})$ one which finds applications in field theory, quantum optics, cosmology and the description of quantum effects in curved space-time.
	\item[-] in Nielsen's approach the counterpart of the choice of universal gates is the choice of right-invariant metric. This is crucial if the ''geodesic'' complexity is considered as providing reasonable estimates on the complexity defined in terms of universal gates. Therefore, it is advantageous to know the structure of geodesics for a wide variety of right-invariant metrics. Apart from reflecting the properties of universal gates it offers deeper insight into the properties of Nielsen's approach. For example, assume we compute the complexity of evolution operator (as a function of time). Time evolution defines a family of automorphisms of the underlying symmetry group resulting in the family of right-invariant metrics. Therefore, in order to compute the complexity of evolution operator it is sufficient to fix one element (to fix time) and compute the lengths of the relevant family of geodesics, terminating at this element. In the particular case of harmonic group, considered here, this can be nicely illustrated by the example of harmonic oscillator perturbed by the external time-dependent force.
\end{itemize}	
	Finally, Nielsen's complexity gives the lower bound on circuit complexity. One can consider various bounds by selecting the families of (properly normalized) metrics (and, possibly, carrying out some averaging). We postpone more detailed discussion of this and related ideas to the coming publication.

\begin{appendices}
\section{}\label{appendA}	
\renewcommand{\theequation}{A\arabic{equation}}
\setcounter{equation}{0}

We collect here some important information concerning the geometry of Lie groups. We found them useful in analyzing the complexity notion. For more details we refer to \cite{Gursey} where the local geometry of Lie groups is discussed.

Let $G$ be a $n$-dimensional Lie group, $\mathcal{G}$ - its Lie algebra spanned by $X_i$'s, $i=1,2,\ldots,n$, obeying 
\begin{equation}
	\com{X_i,X_j}=i\tensor{c}{^k_{ij}}X_k. \label{Ap1}
\end{equation} 
Let $a\equiv (a^1,\ldots,a^n)$ be, in general local, coordinates in the neighbourhood of the unit element $\mathbbm{1}$; one may assume that it corresponds to the origin of coordinates, $g(0)=\mathbbm{1}$. The composition law
\begin{equation}
	g(a)\cdot g(b)=g(c), \label{Ap2}
\end{equation} 
in terms of coordinates read
\begin{equation}
	c^i=\varphi^i(a,b), \label{Ap3}
\end{equation}
with $\varphi (0,a)=a=\varphi(a,0)$, $\varphi(a,\varphi(b,c))=\varphi(\varphi(a,b),c)$. The left- and right-invariant Cartan-Maurer forms read
\begin{align}
	& \omega_L\equiv ig^{-1}\textrm{d}g, \label{Ap4}\\
	& \omega_R\equiv i\textrm{d}gg^{-1}, \label{Ap5}
\end{align}
or, in terms of coordinates
\begin{align}
	& \omega_L=\tensor{\lambda}{_L^j_k}(a)\textrm{d}a^kX_j\equiv\tensor{\omega}{_L^j}X_j, \label{Ap6}\\
		& \omega_R=\tensor{\lambda}{_R^j_k}(a)\textrm{d}a^kX_j\equiv\tensor{\omega}{_R^j}X_j. \label{Ap7}
\end{align}
The coefficients $\tensor{\lambda}{_{L,R}^j_k}(a)$ can be computed from the composition law as follows. One puts: 
 \begin{align}
 	& \tensor{\mu}{_L^i_j}(a)\equiv\frac{\partial\varphi^i(a,b)}{\partial b^j}\bigg\arrowvert_{b=0}, \label{Ap8}\\
 	& \tensor{\mu}{_R^i_j}(a)\equiv\frac{\partial\varphi^i(b,a)}{\partial b^j}\bigg\arrowvert_{b=0}; \label{Ap9}	
 \end{align}
 then $\tensor{\lambda}{_{L,R}^i_j}(a)$ are simply the inverse matrices:
 \begin{align}
 	& \tensor{\lambda}{_L^i_j}(a)\tensor{\mu}{_L^j_k}(a)=\tensor{\delta}{^i_k},\label{Ap10}\\
 		& \tensor{\lambda}{_R^i_j}(a)\tensor{\mu}{_R^j_k}(a)=\tensor{\delta}{^i_k}.\label{Ap11}
 \end{align}
 The matrices $\tensor{\mu}{_{L,R}^i_j}(a)$ define the left- and right-invariant vector fields on $G$:
 \begin{align}
 	& \tensor{\widetilde{X}}{_{L\,i}}\equiv\tensor{\mu}{_L^j_i}(a)\frac{\partial}{\partial a^j},\label{Ap12}\\
 	& \tensor{\widetilde{X}}{_{R\,i}}\equiv\tensor{\mu}{_R^j_i}(a)\frac{\partial}{\partial a^j}.\label{Ap13}
 \end{align}
 Cartan-Maurer forms obey the structure equations:
 \begin{align}
 	& \textrm{d}\omega_L=i\textrm{d}(g^{-1}\textrm{d}g)=-ig^{-1}\textrm{d}g\wedge g^{-1} \textrm{d}g=i\omega_L\wedge\omega_L,\label{Ap14}\\
 	& \textrm{d}\omega_R=i\textrm{d}(g^{-1}\textrm{d}g)=ig^{-1}\textrm{d}g\wedge g^{-1} \textrm{d}g=-i\omega_R\wedge\omega_R, \label{Ap15}
 \end{align}
 which imply
 \begin{align}
 	& \partial_k\tensor{\lambda}{_L^i_j}-\partial_j\tensor{\lambda}{_L^i_k}=-\tensor{c}{^i_{rs}}\tensor{\lambda}{_L^r_k}\tensor{\lambda}{_L^s_j},\label{Ap16}\\
 	& \partial_k\tensor{\lambda}{_R^i_j}-\partial_j\tensor{\lambda}{_R^i_k}=\tensor{c}{^i_{rs}}\tensor{\lambda}{_R^r_k}\tensor{\lambda}{_R^s_j},\label{Ap17}	
 \end{align}
 or, in terms of $\tensor{\mu}{_{L,R}^i_j}$:
 \begin{align}
 	& \tensor{\mu}{_L^k_i}\partial_k\tensor{\mu}{_L^l_j}-\tensor{\mu}{_L^k_j}\partial_k\tensor{\mu}{_L^l_i}=\tensor{c}{^k_{ij}}\tensor{\mu}{_L^l_k}, \label{Ap18}\\
 & \tensor{\mu}{_R^k_i}\partial_k\tensor{\mu}{_R^l_j}-\tensor{\mu}{_R^k_j}\partial_k\tensor{\mu}{_R^l_i}=-\tensor{c}{^k_{ij}}\tensor{\mu}{_R^l_k}. \label{Ap19}	
 \end{align}
 Eqs.~\eqref{Ap18} and \eqref{Ap19} imply the following commutation rules:
 \begin{align}
 	&\com{i\widetilde{X}_{L\,i},i\widetilde{X}_{L\,j}}=i \tensor{c}{^k_{ij}}i\widetilde{X}_{L\,k}, \label{Ap20}\\
 	& 	\com{i\widetilde{X}_{R\,i},i\widetilde{X}_{R\,j}}=-i \tensor{c}{^k_{ij}}i\widetilde{X}_{R\,k}. \label{AP21}
 \end{align}
Consider the adjoint action of $G$ on $\mathcal{G}$:
\begin{equation}
	gX_ig^{-1}=\tensor{D}{^j_i}(g)X_j ;\label{Ap22}
\end{equation}
\eqref{Ap4} and \eqref{Ap5} yield
\begin{equation}
	g\omega_L g^{-1}=\omega_R, \label{Ap23}
\end{equation}
or, by virtue of \eqref{Ap22}:
\begin{align}
	&\tensor{\lambda}{_R^k_j}(a)=\tensor{D}{^k_i}(a)\tensor{\lambda}{_L^i_j}(a),\label{Ap24}\\
&\tensor{\mu}{_L^k_j}(a)=\tensor{D}{^i_j}(a)\tensor{\mu}{_R^k_i}(a).\label{Ap25}
\end{align}
The left- and right-invariant metrics can be defined as follows. One selects a symmetric positive definite matrix $\eta_{ij}$ and put 
\begin{equation}
	\textrm{d}s^2_{L,R}\equiv \eta_{ij}\tensor{\omega}{_{L,R}^i}\tensor{\omega}{_{L,R}^j}=\eta_{ij}\tensor{\lambda}{_{L,R}^i_k}(a)\tensor{\lambda}{_{L,R}^j_l}(a)\textrm{d}a^k\textrm{d}a^l.\label{Ap26}
\end{equation}
If $\eta_{ij}$ is adjoint-invariant then the left and right metrics coincide by virtue of \eqref{Ap24}.

In what follows we consider the right-invariant metrics; the left-invariant case can be dealt with in an analogous manner. The metric tensor, inferred from eq.~\eqref{Ap26}, reads (we omit the subscript $R$):
\begin{align}
& g_{ij}(a)\equiv\eta_{kl}\tensor{\lambda}{_R^k_i}(a)\tensor{\lambda}{_R^l_j}(a)\equiv\eta_{kl}\tensor{\lambda}{^k_i}(a)\tensor{\lambda}{^l_j}(a),\label{Ap27}\\
& g^{ij}(a)=\eta^{kl}\tensor{\mu}{^i_k}(a)\tensor{\mu}{^j_l}(a).\label{Ap28}	
\end{align}	
Eq.~\eqref{Ap27} implies:
\begin{equation}
	\eta_{kl}=g_{ij}(a)\tensor{\mu}{^i_k}(a)\tensor{\mu}{^j_l}(a).\label{Ap29}
\end{equation}
One can conclude from eqs.~\eqref{Ap27}, \eqref{Ap29} that $\tensor{\lambda}{^i_j}(a)$ play the role of moving reper. 

Once the metric tensor is given one can develop the whole machinery of Riemannian geometry. In particular, the Christoffel symbols read:
\begin{equation}
	\tensor{\Gamma}{^i_{jk}}=\frac{1}{2}\tensor{\mu}{^i_l}(\partial_k\tensor{\lambda}{^l_j}+\partial_j\tensor{\lambda}{^l_k})-\frac{1}{2}\tensor{c}{^t_{sr}}\eta_{tm}\eta^{lr}\tensor{\lambda}{^m_j}\tensor{\lambda}{^s_k}\tensor{\mu}{^i_l}-\frac{1}{2}\tensor{c}{^t_{sr}}\eta_{tm}\eta^{lr}\tensor{\lambda}{^m_k}\tensor{\lambda}{^s_j}\tensor{\mu}{^i_l}.\label{Ap30}
\end{equation}
Now, one can write out the geodesic equation:
\begin{equation}
	\frac{\textrm{d}^2a^i}{\textrm{d}s^2}+\tensor{\Gamma}{^i_{jk}}\frac{\textrm{d}a^j}{\textrm{d}s}\frac{\textrm{d}a^k}{\textrm{d}s}=0. \label{Ap31}
\end{equation}
Let us define
\begin{equation}
	\Pi^i\equiv\tensor{\lambda}{^i_j}\frac{\textrm{d}a^j}{\textrm{d}s}\equiv\frac{\omega^i}{\textrm{d}s}.\label{Ap32}
\end{equation}
$\poisson{\Pi^i}$ are nothing but the coordinates of the tangent vector $\frac{\textrm{d}a^i}{\textrm{d}s}$ with respect to the basis spanned by right-invariant vector fields $\widetilde{X}_{R\,i}$. Due to the fact that our metric structure is right-invariant on homogeneous manifold with right $G$-action, one can expect that the coordinate dependence entering \eqref{Ap31} through Christoffel symbols will disappear once this equation is expressed in terms of $\Pi^i$. In fact, we obtain
\begin{equation}
	\frac{\textrm{d}\Pi^i}{\textrm{d}s}=\eta^{ij}\eta_{kl}\tensor{c}{^l_{jm}}\Pi^k\Pi^m. \label{Ap33}
\end{equation}
These are Euler-Arnold equations; they are autonomous nonlinear equations which, when solved, yield $\Pi^i=\Pi^i(s)$.

Once $\Pi^i(s)$ are known one can find $a^i(s)$ by solving first order equations resulting from \eqref{Ap32}:
\begin{equation}
	\frac{\textrm{d}a^i}{\textrm{d}s}=\tensor{\mu}{^i_j}(a)\Pi^j(s). \label{Ap34}
\end{equation}
Eq.~\eqref{Ap31} is derivable from the Lagrangian:
\begin{equation}
	L=\frac{1}{2}g_{ij}(a)\frac{\textrm{d}a^i}{\textrm{d}s}\frac{\textrm{d}a^j}{\textrm{d}s}=\frac{1}{2}\eta_{kl}\tensor{\lambda}{^k_i}(a)\tensor{\lambda}{^l_j}(a)\frac{\textrm{d}a^i}{\textrm{d}s}\frac{\textrm{d}a^j}{\textrm{d}s}.\label{Ap35}
\end{equation}
$L$ defines a nondegenerate Lagrangian system and one can apply all techniques of analytical mechanics. In particular, Noether theorem related to right $G$-action yields $n$ conserved charges. To find their explicit form consider the right shifts
\begin{equation}
	g(a+\delta a)=g(a)g(b), \label{Ap36}
\end{equation}
with $b$ infinitesimal but otherwise arbitrary. Then 
\begin{equation}
	\delta a^i=\tensor{\mu}{_L^i_j}(a)b^j, \label{Ap37}
\end{equation}
and the relevant charges read:
\begin{equation}
	\mathcal{I}_j=\frac{\partial L}{\partial\dot{a}^i}\tensor{\mu}{_L^i_j}(a),\quad j=1,2,\ldots,n.\label{Ap38}
\end{equation}
Note that $\mathcal{I}_j$'s when expressed in terms of phase space variables, do not depend on explicit choice of Lagrangian. This is obvious since we are dealing with Hamiltonian action of $G$ on itself. Eq.~\eqref{Ap18} gives
\begin{equation}
	\poisson{\mathcal{I}_i,\mathcal{I}_j}=-\tensor{c}{^k_{ij}}\mathcal{I}_k. \label{Ap39}
\end{equation}
By virtue of \eqref{Ap25} and \eqref{Ap35} one finds
\begin{equation}
	\mathcal{I}_j=\eta_{kl}\tensor{D}{^k_j}(a)\Pi^l(a).\label{Ap40}
\end{equation}
Eq.~\eqref{Ap40} is also a direct consequence of Euler-Arnold equations \eqref{Ap33} and the definition of $\Pi^i(a)$. To see this let us note that the following identity holds as a consequence of right invariance of $\tensor{\lambda}{_R^i_j}(a)\textrm{d}a^j$, eqs.~\eqref{Ap25} and \eqref{Ap17}:
\begin{equation}
	\partial_k\tensor{D}{^i_j}(a)+\tensor{c}{^i_{rs}}\tensor{D}{^r_j}(a)\tensor{\lambda}{_R^s_k}(a)=0.\label{Ap41}
\end{equation}
Taking the derivative of both sides of \eqref{Ap40} with respect to $s$ and using \eqref{Ap41} and \eqref{Ap33} we conclude that $\mathcal{I}$'s are constant. Once $\Pi^i(s)$ are known (from Euler-Arnold equations) one can find from \eqref{Ap40} $a^i(s)$ provided $\mathcal{I}$'s are functionally independent. However, it is generally not the case (take, for example, $G$ abelian). One has, therefore, address directly to eq.~\eqref{Ap34} or rewrite \eqref{Ap40} as
\begin{equation}
	\frac{\textrm{d}a^i}{\textrm{d}s}=\eta^{jk}{(D^{-1}(a))^l}_k\tensor{\mu}{^i_j}(a)\mathcal{I}_l.\label{Ap42}
\end{equation}
As far as the latter equation is concerned we conclude that the symmetry under right action of $G$ allows us to reduce the set of second order equations \eqref{Ap31} to the one of first order ones, eq.~\eqref{Ap42}. It is, however, more convenient to solve the geodesic equations in two stages. First, we solve Euler-Arnold equations for $\Pi^i(s)$ and then we use eq.~\eqref{Ap34} to find $a^i=a^i(s)$.

Apart from $\mathcal{I}$'s we have still one general (i.e. independent of the particular choice of $\eta_{ij}$) integral of motion - the Hamiltonian, numerically equal to the Lagrangian:
\begin{equation}
	L=\frac{1}{2}\eta^{ij}{(D^{-1}(a))^k}_i{(D^{-1}(a))^l}_j\mathcal{I}_k\mathcal{I}_l.\label{Ap43}
\end{equation}
If $G$ is compact one can choose the basis in the Lie algebra $\mathcal{G}$ such that the structure constant are totally antisymmetric. Then the Euler-Arnold equations can be rewritten in Lax form:  
\begin{equation}
	\frac{\textrm{d}L}{\textrm{d}s}=i\com{L,M}, \label{Ap44}
\end{equation}
with
\begin{align}
	&L\equiv\eta_{ij}\Pi^iX_j,\label{Ap45}\\
	& M\equiv \Pi^iX_i.\label{Ap46} 
\end{align}
The integrals of motion are obtained from the traces of the powers of $L$:
\begin{equation}
	\textrm{Tr}(L^k)={(D^{-1}(a))^{l_1}}_{j_1}\dots{(D^{-1}(a))^{l_k}}_{j_k}\textrm{Tr}(X_{j_1}\dots X_{j_k})\mathcal{I}_{l_1}\dots\mathcal{I}_{l_k}.\label{Ap47}
\end{equation}
In the basis adopted the matrices $D(a)$ are orthogonal while the traces are ad-invariant. Therefore, we finally find:
\begin{equation}
	\textrm{Tr}(L^k)=d_{l_1\dots l_k}\mathcal{I}_{l_1}\dots\mathcal{I}_{l_k},\label{Ap48}
\end{equation}
so no new integrals of motion are obtained. 

Finally, we find the conditions for the exponential mapping to define the geodesic. Let
\begin{equation}
	X\equiv x^iX_i, \label{Ap49}
\end{equation}
and
\begin{equation}
	g(t)=e^{itX}.\label{Ap50}
\end{equation}
In terms of coordinates, $g(t)=g(a(t))$, eq.~\eqref{Ap50} can be written as
\begin{equation}
\frac{\textrm{d}a^i}{\textrm{d}t}=\tensor{\mu}{_R^i_j}(a)x^j, \label{Ap51}
\end{equation} 
or
\begin{equation}
	\tensor{\lambda}{_R^i_j}(a)\textrm{d}a^j=x^k\textrm{d}t ,\label{Ap52}
\end{equation}
and 
\begin{equation}
	\textrm{d}s^2=\eta_{kl}x^kx^l\textrm{d}t^2.\label{Ap53}
\end{equation}
Therefore, after suitable rescaling of $x_i$'s one can use $s$ instead of $t$. Then \eqref{Ap32} implies:
\begin{equation}
	\Pi^i(s)=x^i, \label{Ap54}
\end{equation}
and \eqref{Ap33} is obeyed provided
\begin{equation}
	\eta_{kl}\tensor{c}{^l_{jm}}x^kx^m=0. \label{Ap55}
\end{equation}

\end{appendices}

	\end{document}